\documentclass[apj]{emulateapj}
\usepackage{apjfonts} 
\usepackage{graphicx}

\def\gtrsim{\mathrel{\hbox{\rlap{\hbox{\lower4pt\hbox{$\sim$}}}\hbox{$>$}}}}
\def\lesssim{\mathrel{\hbox{\rlap{\hbox{\lower4pt\hbox{$\sim$}}}\hbox{$<$}}}}
\def\gtrsim{\mathrel{\hbox{\rlap{\hbox{\lower4pt\hbox{$\sim$}}}\hbox{$>$}}}}
\def\lesssim{\mathrel{\hbox{\rlap{\hbox{\lower4pt\hbox{$\sim$}}}\hbox{$<$}}}}

\received{September 11, 2009}
\accepted{September 24, 2009}
\journalinfo{Astrophysical~journal}
\submitted{}

\shortauthors{Homan et al.}
\shorttitle{MOJAVE. VII. Blazar Jet Acceleration}

\begin{document}
\title{MOJAVE: Monitoring of Jets in AGN with VLBA Experiments. VII. Blazar Jet Acceleration}
\author{
D. C. Homan\altaffilmark{1},
M. Kadler\altaffilmark{2,3,4,5},
K. I. Kellermann\altaffilmark{6},
Y. Y. Kovalev\altaffilmark{7,8},
M. L. Lister\altaffilmark{9},
E. Ros\altaffilmark{10,7},
T. Savolainen\altaffilmark{7},
J.~A.~Zensus\altaffilmark{7}
}
\altaffiltext{1}{
Department of Physics and Astronomy, Denison University,
Granville, OH 43023;
\email{homand@denison.edu}
}
\altaffiltext{2}{
Dr.\ Remeis-Sternwarte Bamberg, Universit\"at Erlangen-N\"urnberg,
Sternwartstrasse 7, 96049 Bamberg, Germany;
\email{matthias.kadler@sternwarte.uni-erlangen.de}
}
\altaffiltext{3}{
Erlangen Centre for Astroparticle Physics, Erwin-Rommel Str.~1,
91058 Erlangen, Germany
}
\altaffiltext{4}{
CRESST/NASA Goddard Space Flight Center, Greenbelt, MD 20771, USA
}
\altaffiltext{5}{
Universities Space Research Association, 10211
Wincopin Circle, Suite 500 Columbia, MD 21044, USA
}
\altaffiltext{6}{
National Radio Astronomy Observatory, 520 Edgemont Road,
Charlottesville, VA 22903-2475; 
\email{kkellerm@nrao.edu}
}
\altaffiltext{7}{
Max-Planck-Institut f\"ur Radioastronomie, Auf dem H\"ugel 69,
D-53121 Bonn, Germany;
\email{tsavolainen@mpifr-bonn.mpg.de, azensus@mpifr-bonn.mpg.de}
}
\altaffiltext{8}{
Astro Space Center of Lebedev Physical Institute,
Profsoyuznaya 84/32, 117997 Moscow, Russia; 
\email{yyk@asc.rssi.ru}
}
\altaffiltext{9}{
Department of Physics, Purdue University, 525 Northwestern
Avenue, West Lafayette, IN 47907;
\email{mlister@purdue.edu}
}
\altaffiltext{10}{
Departament d'Astronomia i Astrof\'{\i}sica, Universitat de Val\`encia,
E-46100 Burjassot, Val\`encia, Spain;
\email{Eduardo.Ros@uv.es}
}

\begin{abstract}
We discuss acceleration measurements for a large sample of extragalactic
radio jets from the MOJAVE program which studies the parsec-scale jet
structure and kinematics of a complete, flux-density-limited sample of
Active Galactic Nuclei (AGN). Accelerations are measured from the
apparent motion of individual jet features or ``components'' which may
represent patterns in the jet flow. We find that significant
accelerations are common both parallel and perpendicular to the observed
component velocities. Parallel accelerations, representing changes in
apparent speed, are generally larger than perpendicular acceleration
that represent changes in apparent direction. The trend for larger
parallel accelerations indicates that a significant fraction of these
changes in apparent speed are due to changes in intrinsic speed of the
component rather than changes in direction to the line of sight. We find
an overall tendency for components with increasing apparent speed to be
closer to the base of their jets than components with decreasing
apparent speed. This suggests a link between the observed pattern
motions and the underlying flow which, in some cases, may increase in
speed close to the base and decrease in speed further out; however,
common hydro-dynamical processes for propagating shocks may also play a
role. About half of the components show ``non-radial'' motion, or a
misalignment between the component's structural position angle and its
velocity direction, and these misalignments generally better align the
component motion with the downstream emission.  Perpendicular
accelerations are closely linked with non-radial motion.  When observed
together, perpendicular accelerations are usually in the correct
direction to have caused the observed misalignment.
\end{abstract}


\section{Introduction}

The acceleration and collimation of powerful extragalactic jets
associated with some Active Galactic Nuclei (AGN) is still not fully
understood. While it seems clear that strong magnetic fields associated
with the supermassive black hole/accretion disk system play a key role
in the initial acceleration and collimation of the jet
\citep[e.g.][]{Meier01}, it is unclear whether this process is largely
complete with the conversion of Poynting flux to flow energy on
subparsec scales \citep[e.g.][]{S05} or continues into the parsec and
decaparsec scales observed with Very Long Baseline Interferometry
(VLBI). If jets are still being accelerated and collimated on these
longer length scales, is the process largely hydrodynamic in nature, or
is there still a significant, perhaps dominant, magnetic contribution as
suggested by \citet{VK04}?

To date, the observational evidence from VLBI observations of blazars
jets on parsec to decaparsec scales presents a mixed picture. Detailed
studies of individual jets have reported changes in speed and trajectory
of particular components (bright jet features) as well as speed
differences between multiple components in the same jet
\citep[e.g.][]{WCRB94,Z95,W01,G01,H03,B05,SWVT06}; however, taken
together there is no clear trend for how, or if, jets accelerate on
parsec-scales. In a larger study of gamma-ray blazars \citet{J05} 
reported accelerations in nine of the fifteen jets they studied, and
these tended to be positive accelerations, indicating increased apparent
speed, although they could not determine whether the observed
accelerations were due to changes in intrinsic speed or direction of the
components. Other authors of multi-jet studies have reported a tendency
for more distant jet components to have faster speeds than nearby
components in the same object \citep{H01,P06,B08}; however, \citet{H01}
also looked for accelerations in the motions of individual jet
components and did not find evidence for large accelerations. 
\citet{KL04} pointed out that higher frequency VLBI studies tend to
observe systematically faster speeds than lower frequency studies which
probe longer length scales, suggesting that components are instead
faster at smaller jet separations. 

Large apparent opening angles and variations in jet ejection angles are
common in VLBI observations of blazar jets
\citep[e.g][]{SCS03,JML04,J05,L06}; however, jets observed on kiloparsec
scales tend to be well focused with narrow opening angles.  The high
incidence of ``non-radial'' or non-ballistic motion observed on VLBI
scales \citep{KL04,PMF07} suggest that jet components can change
direction after ejection from the core, and \citet{KL04} showed that
these non-radial motions tend to be in the direction of the downstream
flow. Whether these non-radial motions represent evidence for
collimation or simply bends in already well collimated jets is unclear;
however \citet{H03} did observe a large, sudden ($\simeq 25^\circ$)
change in trajectory in a powerful jet component in 3C\,279 to better
align with the direction and speed of other jet components and the
arcsecond jet direction. Given the wide variety of ejection angles and
apparent speeds in this jet \citep[e.g.][]{W01}, Homan et al. suggested
this event was evidence that the collimation process may continue even
to kiloparsec scales (deprojected). \citet{JML04} observed a similar
change in another jet component in this same jet, but much closer to the
base of the jet.

Here we report the results of the first large scale study of
accelerations in the apparent motions of blazar jets. We examine
accelerations both parallel and perpendicular to the apparent component
velocities, representing changes in apparent speed and direction
respectively, and we analyze these acceleration measurements for 203 of
the best determined jet component motions from the MOJAVE program
\citep{LH05,LAA09}. This work is an extension of our kinematical
analysis of the MOJAVE~I sample (\citealt{LCH09},
hereafter Paper~VI). MOJAVE stands for Monitoring Of Jets in Active
galaxies with VLBA Experiments, and it is an ongoing program using the
National Radio Astronomy Observatory's Very Long Baseline Array (VLBA)
to monitor the parsec-scale structure and polarization of the brightest
AGN jets at $\lambda 2$ cm. MOJAVE builds upon the older {{\em 2cm
Survey} \citep{K98,Z02,KL04,K05}, and thus, in some cases, our time
baseline for kinematical study of jet components spans more than a
decade. Previous publications from the MOJAVE program have studied the
parsec-scale polarization properties of the MOJAVE I sample
\citep{LH05,HL06}, their VLA structure \citep{CLK07}, their parent
luminosity function \citep{CL08}, and the connection with their
gamma-ray properties measured by the \textit{Fermi} satellite 
\citep[][Savolainen et al, submitted to Science]{K09,LHK09}.

This paper is organized as follows. Section 2 summarizes key
relationships describing the connection between observed accelerations
and changes in intrinsic properties of the component or pattern motion,
such as Lorentz factor and angle to the line of sight. Section 3
describes the sample of jet components studied for acceleration and
reports the results of our analysis for trends in the observed
accelerations. Section 4 discusses our results and a summary of our 
conclusions appears in \S{5}. The appendix provides additional
background for \S{2}. 

\section{Relationships for Accelerating Motion} 

In this section we summarize relationships describing changes in
apparent speed of parsec-scale AGN jet features. These expressions are
derived in more detail in the appendix and build upon the relationships
laid out in \citet{BK79} and sources referenced therein. 

A component or pattern moving with an intrinsic speed $\beta c$ at angle
$\theta$ to the line of sight will exhibit an observed motion given by
the familiar form
\begin{equation}
\label{e:beta_obs}
\beta_{obs} = \frac{\beta\sin\theta}{1-\beta\cos\theta}.
\end{equation}
In practice, we deduce $\beta_{obs}$ from the observed angular speed,
$\mu$, and redshift, $z$ from
\begin{equation}
\beta_{obs} = \mu D_A (1+z)
\end{equation}
where $D_A$ is the angular size distance to the galaxy.

As illustrated in the appendix, the intrinsic velocity, $\vec{\beta}$,
is actually a vector which can change in speed and/or direction,
resulting in apparent changes to $\vec{\beta}_{obs}$ in the plane of the
sky. We characterize these changes as either {\em parallel} or {\em
perpendicular} to $\vec{\beta}_{obs}$ or equivalently the observed
angular velocity $\vec{\mu}$. 

\subsection{Apparent Changes Along Component Motion}
\label{s:parallel}

Changes along the observed motion, or {\em parallel} accelerations, 
are given by the following expression (Appendix A.2):
\begin{equation}
\label{e:dbeta_obs1}
\frac{d\beta_{\parallel obs}}{dt_{obs}} = \frac{\dot{\beta}\sin\theta+\beta\dot{\theta}(\cos\theta-\beta)}{(1-\beta\cos\theta)^3}
\end{equation}
where $\dot{\beta}$ and $\dot{\theta}$ are the intrinsic rates of change
of the component speed and angle to the line of sight respectively. A
key question in the evaluation of our observational results is the 
extent to which we can attribute any observed parallel accelerations to
changes in the intrinsic component speed, $\beta$, and in section
\ref{s:accel_ratio} we look at the ratio of parallel to perpendicular
accelerations as a possible diagnostic. 

If observed parallel accelerations are due entirely to changes in the
component's intrinsic speed, $\beta$, or equivalently Lorentz factor,
$\Gamma = 1/\sqrt{1-\beta^2}$, we find the simple relation:
\begin{equation}
\frac{\dot{\beta}_{\parallel obs}}{\beta_{obs}} = \frac{\dot{\mu}_\parallel}{\mu}(1+z) 
= \frac{\dot{\Gamma}}{\Gamma}\frac{\delta^2}{\beta^2} .
\end{equation}
where we have taken the ratio with the apparent speed
$\beta_{obs}$ and substituted the Doppler factor for the component
motion: $\delta = 1/(\Gamma(1-\beta\cos\theta))$. Taking the ratio with
apparent speed provides a convenient measure that can be easily compared
between jets and constructed directly from the observed angular motion.

Hence, under the assumption that all of the observed parallel
acceleration is due to intrinsic speed changes of the component or
pattern, the ratio of observed angular acceleration to observed angular
speed gives a fairly simple result in terms of intrinsic changes in the
component's Lorentz factor. The $\delta^2$ term implies that small rates
of change in intrinsic Lorentz factor should be greatly magnified in the
observed angular acceleration for the high $\delta$ jets typical of the
MOJAVE I sample (e.g. Paper VI).

\subsection{Apparent Changes Perpendicular to the Component Motion}
\label{s:accel_ratio}

``Perpendicular'' accelerations, or changes in motion transverse to the
apparent motion on the plane of the sky, are given by the following
expression (Appendix A2):
\begin{equation}
\label{e:dbeta_obs2}
\frac{d\beta_{\perp obs}}{dt_{obs}} =
\frac{\beta\dot{\phi}\sin\theta}{(1-\beta\cos\theta)^2},
\end{equation}
where $\phi$ is the azimuthal angle of both the component's intrinsic 
velocity, $\vec{\beta}$, and its apparent velocity in the plane of the
sky, $\vec{\beta}_{obs}$. See the appendix for an explanation of the
geometry of $\vec{\beta}$.

The relative magnitudes of observed parallel and perpendicular
accelerations may give us information about how much of any observed
parallel acceleration is due to intrinsic changes in component speed as
opposed to changes in its angle to the line of sight. If all the
apparent acceleration is due only to changes in direction of the
component motion, we can divide eqn. \ref{e:dbeta_obs1} by eqn.
\ref{e:dbeta_obs2} and set $\dot{\beta} = 0$ to get the following
expression:
\begin{equation}
\frac{d\beta_{\parallel obs}/dt_{obs}}{d\beta_{\perp obs}/dt_{obs}}=\frac{\dot{\beta}_\theta}{\dot{\beta}_\phi}
\frac{\cos\theta-\beta}{1-\beta\cos\theta} = \frac{\dot{\beta}_\theta}{\dot{\beta}_\phi}\cos\theta_{ab}.
\end{equation}
Here $\theta_{ab}$ is the aberrated angle to the line of sight
in the frame co-moving with the jet component and we have defined
$\dot{\beta}_\theta = \beta\dot{\theta}$ and $\dot{\beta}_\phi
=\beta\dot{\phi}\sin\theta$. By symmetry, we expect that
$\dot{\beta}_\theta$ and $\dot{\beta}_\phi$, the components of
directional change of $\vec{\beta}$, will be of similar magnitude when
averaged over many jets. For components moving at the critical angle for
maximal superluminal motion, $\cos\theta=\beta$, $\cos\theta_{ab} = 0$,
and small changes in $\theta$ will not change the apparent speed,
$\beta_{obs}$; thus only perpendicular acceleration is possible from a
change in direction. A typical beamed jet in a flux-density-limited
survey like MOJAVE has an angle to the line of sight $\theta\sim
0.5\Gamma^{-1}$ \citep{LM97} giving $\cos\theta_{ab} \simeq 0.6$. 
Therefore, if all of the observed accelerations in our sample are due to
changes in direction rather than intrinsic changes in speed or Lorentz
factor, we expect the observed parallel accelerations to be about 60\%
of the magnitude of perpendicular accelerations when averaged across our
sample.

\section{Data and Results}
\label{s:results}

In Paper VI we reported proper motion results for the MOJAVE I complete
sample of 135 parsec-scale radio jets. Vector proper motions were fit to
position versus time data for 526 robust components in 127 of the 135
jets. In each jet, component positions were measured relative to the jet
base or ``core'' which was assumed to be stationary from epoch to epoch.
For the 311 components with 10 or more epochs of observation,  constant
acceleration terms were included in the apparent proper motion fit,
resulting in measurement of the apparent parallel,
$\dot{\mu}_\parallel$, and perpendicular, $\dot{\mu}_\perp$,
acceleration. These parallel and perpendicular directions are taken
relative to the average vector proper motion on the sky and thus
represent changes in apparent speed and direction respectively. As
described in Paper VI, these accelerations are measured relative to the
middle epoch, and they represent constant apparent acceleration applied
only during the observed epochs. We do not know if the same or different
accelerations were applied to the component motion prior to our
observations. Procedural details and full results of the proper motion
fitting are presented in Paper VI, and here we discuss a subset of
components with the highest quality proper motion measurements to
investigate the properties of non-radial and accelerated motion 
observed in our sample. 

For this analysis we restrict our sample of jet motions to those having
a proper motion, $\mu$, of at least $3\sigma$ significance, $\geq 10$
epochs (required for our acceleration analysis), a known redshift
(required for relative acceleration analysis of
$\dot{\beta}_{obs}/\beta_{obs}$), and an uncertainty in misalignment
angle, $|\langle\vartheta\rangle-\phi|$, of no more than $5^\circ$ 
between the average component position angle, $\langle\vartheta\rangle$,
and vector proper motion position angle, $\phi$. This limit guarantees
that both angles in this quantity are well determined, and therefore
that the meaning of the parallel and perpendicular accelerations,
defined relative to $\phi$, are unambiguous. A total of 203 jet
components meet these criteria, and these components are listed in Table
\ref{t:comps} along with selected properties from the proper motion
results. 

Plots illustrating examples of accelerated motion in a few of the
components in our sample are briefly discussed in the following
subsection. Separation versus time and sky position plots for all the
203 jet components analyzed in this paper, along with the entire MOJAVE
sample, are included in Paper VI. In the remainder of this section, we
present results on the general properties of non-radial motions and
accelerations, and we specifically investigate a number of key
relationships including: (1) the connection between non-radial motion
and downstream jet structure, (2) the relative magnitudes of apparent 
parallel and perpendicular accelerations which are either along or
transverse to the component motion respectively, (3) the magnitude and
sign of accelerations as a function of distance along the jet, and (4)
the relationship between observed accelerations and non-radial motion. 

\subsection{Examples of Accelerated Motion}

\begin{figure*}[t]
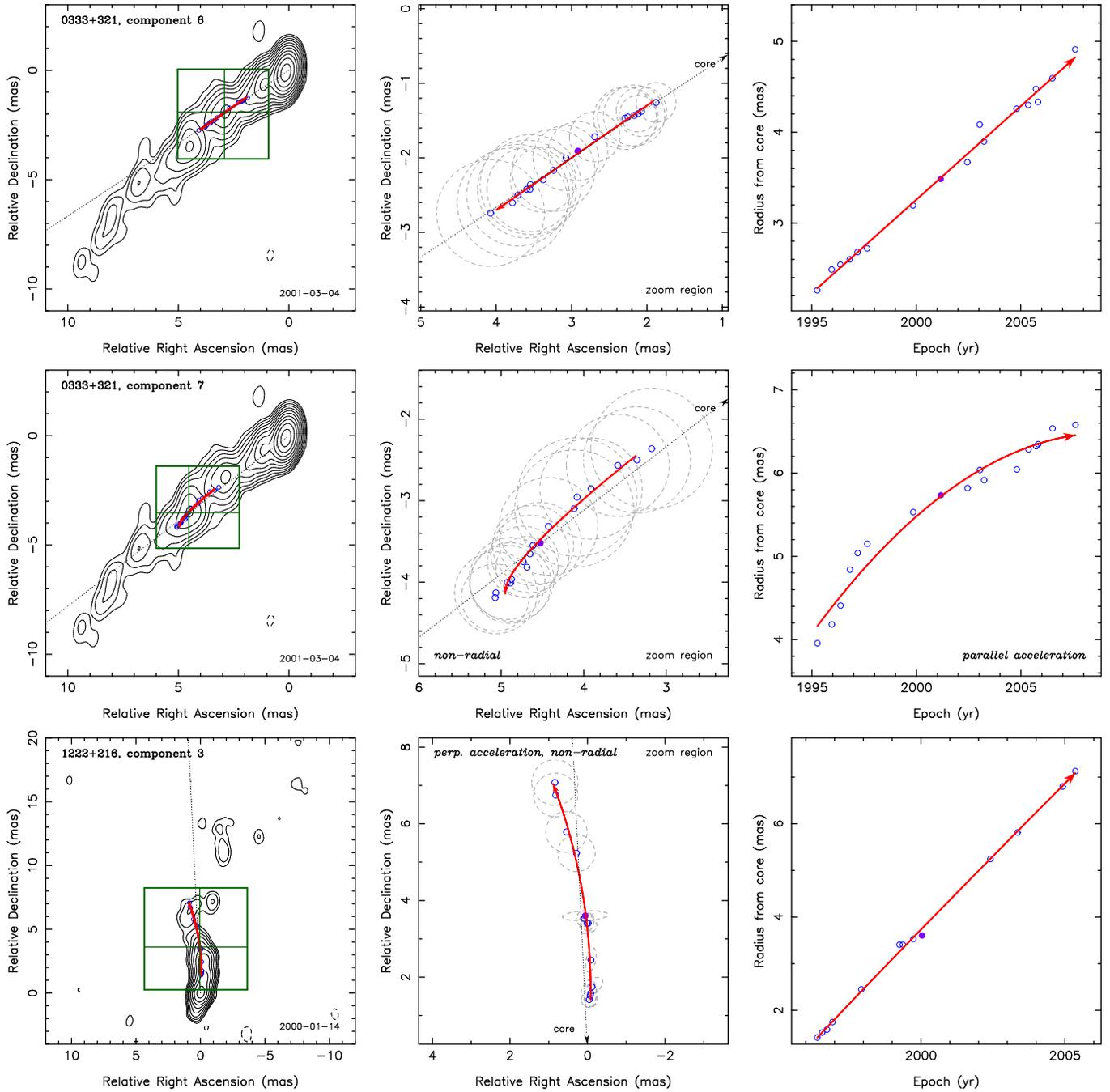

\centering
\resizebox{1.0\hsize}{!}{\includegraphics[angle=-90,trim=0cm 0cm -0.2cm 0cm]{fig1a.eps}}
\resizebox{1.0\hsize}{!}{\includegraphics[angle=-90,trim=0cm 0cm -0.05cm 0cm]{fig1b.eps}}
\resizebox{1.0\hsize}{!}{\includegraphics[angle=-90,trim=0cm 0cm -0.1cm 0cm]{fig1c.eps}}
\caption{
\label{f:examples1}
Plots of component trajectories in 0333$+$321 and 1222$+$216. One
component appears per row. Left hand panels contain a contour image of
the jet for an epoch near the midpoint of the observations of the
component. Superimposed on the contour are the sky position of the
component for all epochs, a radial line from the jet core to the mean
component position, and a projection of our best fit trajectory for the
component. Middle panels show a zoomed view of the component's sky
positions, radial line, and fitted trajectory.
The dotted outlines represent the full-width half-maximum size of
the component in each epoch.
Right hand panels show
the radial distance of the component from the jet core as a function of
time with a projection of our best fit trajectory. 
}
\end{figure*}

\begin{figure*}[t]
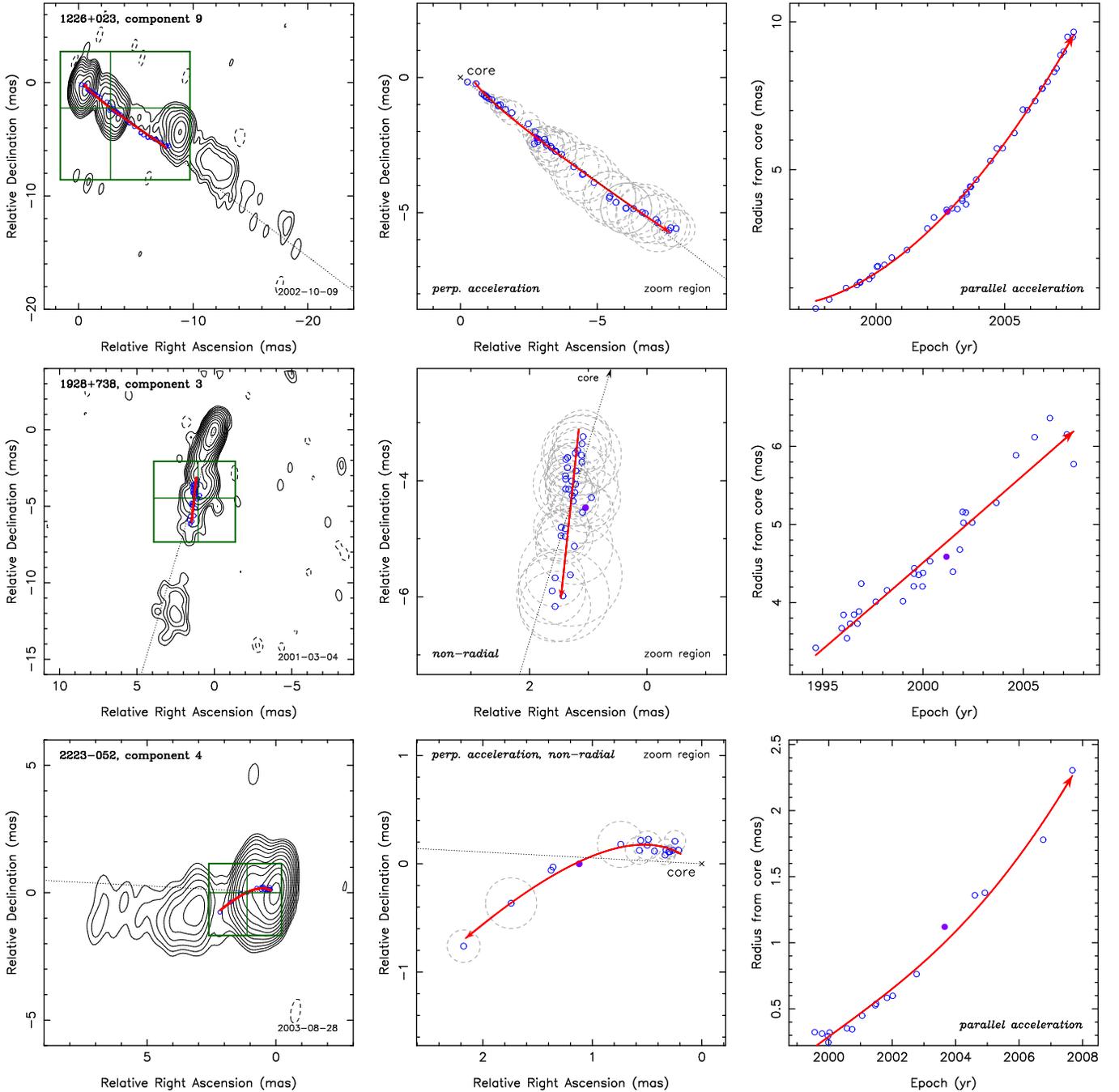

\centering
\resizebox{1.0\hsize}{!}{\includegraphics[angle=-90,trim=0cm 0cm -0.2cm 0cm]{fig2a.eps}}
\resizebox{1.0\hsize}{!}{\includegraphics[angle=-90,trim=0cm 0cm -0.2cm 0cm]{fig2b.eps}}
\resizebox{1.0\hsize}{!}{\includegraphics[angle=-90,trim=0cm 0cm -0.1cm 0cm]{fig2c.eps}}
\caption{
\label{f:examples2}
Plots of component trajectories in 3C\,273, 1928$+$738, and 2223$-$052,
otherwise as described for Figure~\ref{f:examples1}.
}
\end{figure*}

Figures~\ref{f:examples1},~\ref{f:examples2} illustrate examples of the types
of motion observed in the components in our sample. Six components are
illustrated, three per figure. Each row of the figures shows the motion
of one component in three ways: first superimposed on the jet structure
at a middle epoch of our observations, second as a zoomed-in view to
better show the motion on the sky, and third as a radial distance versus
time plot to illustrate the speed of the component as a function of
time. Each panel also contains a colored solid line indicating our best
fit to the component position versus time including acceleration terms.

The first two rows of Figure~\ref{f:examples1} show components 6 and 7
of 0333$+$321. Component 6 is an example of a component which is moving
radially outward at a constant speed and doesn't show signs of
acceleration. Component 7 is further out in the jet and is moving
non-radially in the direction of the downstream flow. Component 7 is
also showing a significant negative parallel acceleration, i.e. it is
slowing down during our observations. While the trajectory of component
7 in the middle panel appears somewhat curved, the measured
perpendicular acceleration is not significant.

The last row of Figure~\ref{f:examples1} shows the motion of component 3 from
1222$+$216. The trajectory is clearly curved to the east as the
component follows the bent path of the jet, and the measured
perpendicular acceleration corresponding to this curvature is highly
significant. There is no accompanying change in the apparent speed of
the component as indicated by the lack of significant parallel
acceleration. It is interesting that four jet components in 1222$+$216
show essentially the same motion with significant perpendicular
accelerations producing curving motion to the east, indicating a stable
curved path for the jet out to about 10 milli-arcseconds from the jet
core. It is interesting that this 'stable' path is not directed towards
the most distant jet component, component 1, which seems to be following
its own, roughly parallel trajectory, at about half the apparent speed
of the other features.

Some jets, like 1222$+$216, show consistent patterns of accelerated 
motion for several components, while others, like 3C\,273 (1226$+$023),
show a wide variety of accelerated or non-radial motion types. The first
row of Figure~\ref{f:examples2} shows the motion of component 9 of
3C\,273. Component 9 is an example of a component with radially outward
motion that is increasing in apparent speed, i.e. it is showing a
significant positive parallel acceleration. The component also has a
significant perpendicular acceleration, tending to curve the trajectory
upward slightly, but this perpendicular acceleration is much smaller
than the parallel acceleration producing the observed increase in
apparent speed.

Another jet which shows a wide variety of accelerated and non-radial
motion types is 1928$+$738. The component illustrated in the second row
of Figure~\ref{f:examples2}, component 3, is an example of a jet component that shows
more jitter than is typical in its centroid position from epoch to
epoch; however, despite this limitation, our large number of epochs
still allow us to determine its mean trajectory very well. We find the
component to be moving non-radially toward the downstream jet structure,
and we find no significant acceleration either parallel or perpendicular
to the component motion.

Finally, the last row in Figure~\ref{f:examples2} shows the motion of
component 4 in 2223$-$052. This component is an example of non-radial
motion with significant acceleration both parallel to the component
velocity, resulting in an increasing apparent speed, and perpendicular
to the component velocity, resulting in a trajectory curving toward the
downstream jet structure.

\subsection{Non-radial Motion}

We characterize the extent to which a jet component's motion is
non-radial by its misalignment angle, $|\langle\vartheta\rangle-\phi|$. 
Figure~\ref{f:nonrad_hist} is a histogram of these values for the 203
jet components in our sample. Approximately one half of the components
($N = 100$) are significantly non-radial with 
$|\langle\vartheta\rangle-\phi| > 0$ at the $\geq 3\sigma$ level, and
about a fifth ($N=41$) have large non-radial motion with 
$|\langle\vartheta\rangle-\phi| \gtrsim 10$ degrees by at least
$2\sigma$. 

\begin{figure}[t]
\centering
\resizebox{1.0\hsize}{!}{\includegraphics[angle=-90]{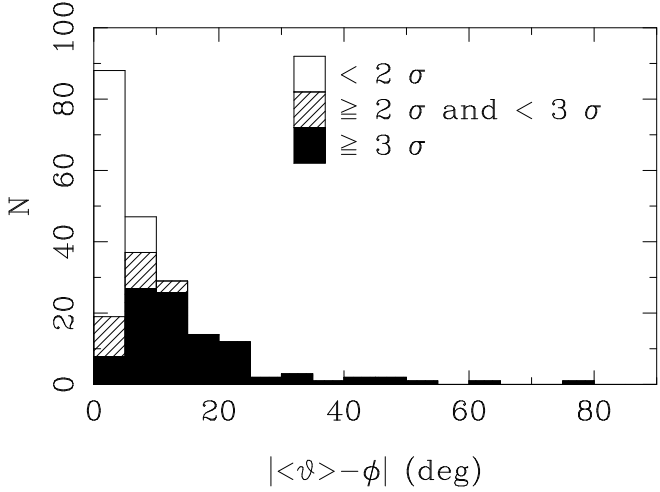}}
\caption{
\label{f:nonrad_hist}
Distribution of velocity vector alignment parameter, 
$|\langle\vartheta\rangle-\phi|$, for the 203 jet components in our 
restricted sample. Hash and solid fill styles indicate non-radial
motion, $|\langle\vartheta\rangle-\phi| > 0$, significant at the
$2-3\sigma$ and $\geq 3\sigma$ levels respectively.
}
\end{figure}

For each of the significantly non-radial components, we compared the
direction of non-radial motion to a contour image of the jet from an
epoch close to the middle epoch of that component's trajectory. We found
that the large majority of non-radial components, 69 of the 83 for which
we could make a determination, are moving in a direction, relative to
their radial position angle, which will tend to make them better aligned
with the downstream emission. In cases where the downstream position
angle was ambiguous due to a curving jet, we took the next structure in
the jet to define the downstream direction. In a few cases, such as
3C\,279 component 1, we were able to use the direction of the extended
structure from lower-resolution observations to make this comparison.

\subsection{Acceleration}
\label{s:accel_results}

Of the 203 jet components in our sample, approximately one third of ($N
= 64$) have significant parallel acceleration and one fifth ($N = 44$) 
have significant perpendicular acceleration with $\dot{\mu}_\parallel$
or $\dot{\mu}_\perp$ respectively differing from zero at the $\geq
3\sigma$ level.

To compare accelerations in jet components with different apparent
speeds, we define relative accelerations as follows:
\begin{equation}
\dot{\eta}_\parallel = \dot{\beta}_{\parallel obs}/\beta_{obs} = (1+z)\dot{\mu}_\parallel/\mu\;,
\end{equation}
and
\begin{equation}
\dot{\eta}_\perp = \dot{\beta}_{\perp obs}/\beta_{obs} = (1+z)\dot{\mu}_\perp/\mu\;.
\end{equation}
These relative accelerations are computed for each component in our
sample, and they are listed in Table \ref{t:comps}. We define components
with ``high'' acceleration as those with relative accelerations which
are robustly larger than $0.1$, corresponding to a $10$\% change in the
apparent vector velocity per year in our frame. We find that fifty-one
components have high parallel accelerations and twenty-nine have high
perpendicular accelerations with magnitudes that are at least $2\sigma$
above $0.1$. 

\begin{figure}[t]
\centering
\resizebox{0.93\hsize}{!}{\includegraphics[angle=0]{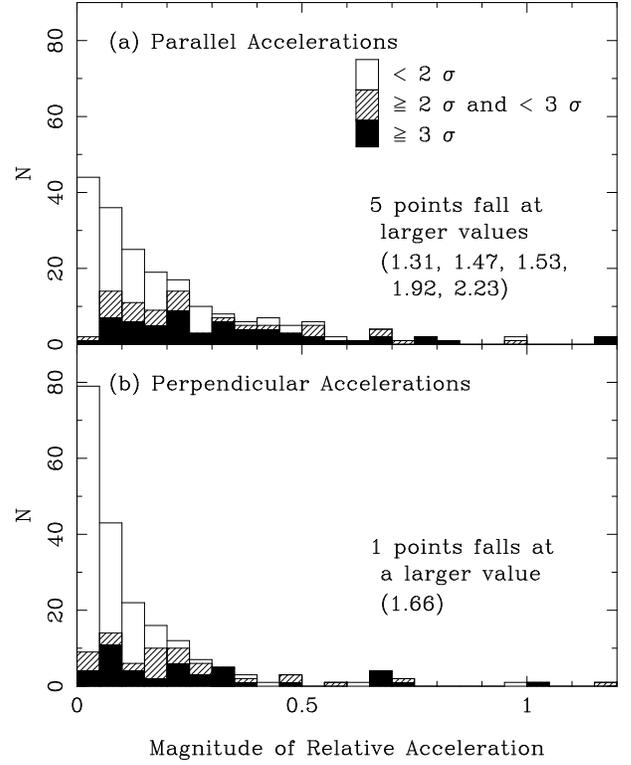}}
\caption{
\label{f:accel_hist}
Histograms of magnitudes of relative accelerations, parallel,
$\dot{\eta}_\parallel$ (panel (a)), and perpendicular,
$\dot{\eta}_\perp$ (panel (b)), to the apparent component motion. Hash
and solid fill styles indicated angular acceleration significant at the
$2-3\sigma$ and $\geq 3\sigma$ levels respectively. As indicated, five
parallel accelerations and one perpendicular acceleration lie to the
right of the plot.
}
\end{figure}

\begin{figure}[b]
\centering
\resizebox{0.93\hsize}{!}{\includegraphics[angle=0]{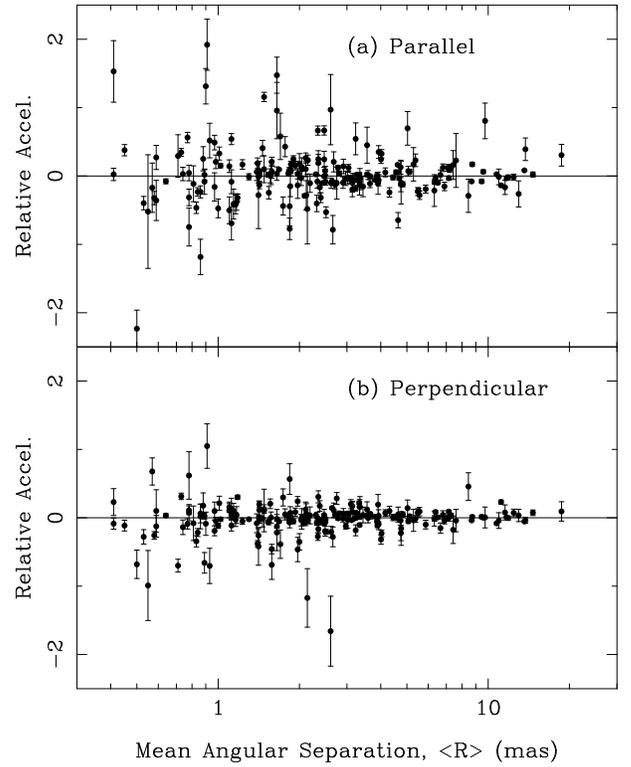}}
\caption{
\label{f:accel_v_R}
Relative accelerations, parallel, $\dot{\eta}_\parallel$ (panel (a)),
and perpendicular, $\dot{\eta}_\perp$ (panel (b)), to the apparent
component motion plotted against average radial distance of the
component from the core position in milli-arcseconds.
}
\end{figure}

Figure~\ref{f:accel_hist} shows histograms of the magnitudes of all the
relative acceleration measurements parallel, $\dot{\eta}_\parallel$, and
perpendicular, $\dot{\eta}_\perp$ to the apparent component motion. The
distribution of relative parallel accelerations is much broader than the
distribution of relative perpendicular accelerations, and a
Kolmogorov-Smirnov (K-S) test gives a probability of only $P<0.001$ that
they are drawn from the same distribution. The relative parallel
accelerations have an average magnitude of $0.25$ while the
perpendicular accelerations average only $0.15$. We find that this
difference between parallel and perpendicular accelerations remains
highly significant whether we divide our sample to only look at nearby
sources with redshifts of $z < 0.1$ or only more distant sources with $z
> 0.1$. These differences are also highly significant whether or not we
include components at mean angular separations $\langle R\rangle < 2.0$
milli-arcseconds, which appear to have less well determined
accelerations (see below).

One possible complicating factor is that there might be a greater
uncertainty in parallel accelerations as those accelerations are taken
along the component motion which tends to be along the jet direction. 
Along the jet direction there is greater opportunity for confusion from
nearby components just up or downstream as well as jitter in the
measured core position along the jet direction due to the emergence of
new features. We find a marginally significant difference between the
distributions of uncertainties for the parallel and perpendicular 
accelerations with a K-S test giving a probability $P = 0.055$ that 
they are drawn from the same distribution. However, this marginal
difference in uncertainties between the parallel and perpendicular
accelerations does not appear to drive the difference in the magnitudes
of the measured accelerations. Indeed, if we only examine components
with uncertainties in their parallel and perpendicular accelerations
which differ by less than 10\%, we find that the difference in magnitude
between the parallel and perpendicular accelerations remains significant
with $P=0.013$ and mean values of $0.23$ and $0.15$ respectively.

\begin{figure}[t]
\centering
\resizebox{1.0\hsize}{!}{\includegraphics[angle=0]{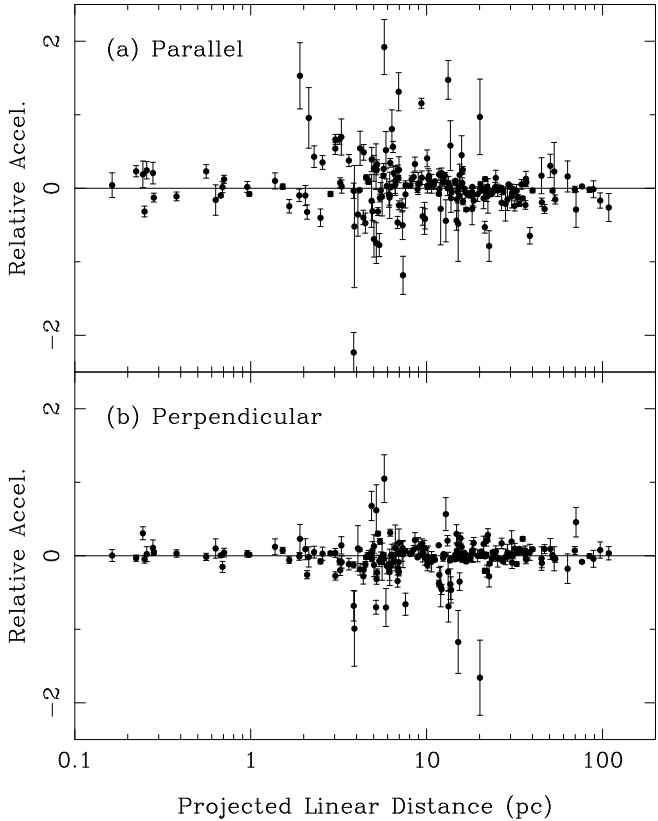}}
\caption{
\label{f:accel_v_pc}
Relative accelerations, parallel, $\dot{\eta}_\parallel$ (panel (a)), and perpendicular,
$\dot{\eta}_\perp$ (panel (b)), to the apparent component motion plotted against average projected
linear distance of the component from the core position in parsecs.  All components with an
projected linear distance of $\leq 1.0$ pc are from 0238$-$084 (NGC 1052) at $z = 0.005$.
}
\end{figure}

\begin{figure}[t]
\centering
\resizebox{1.0\hsize}{!}{\includegraphics[angle=0]{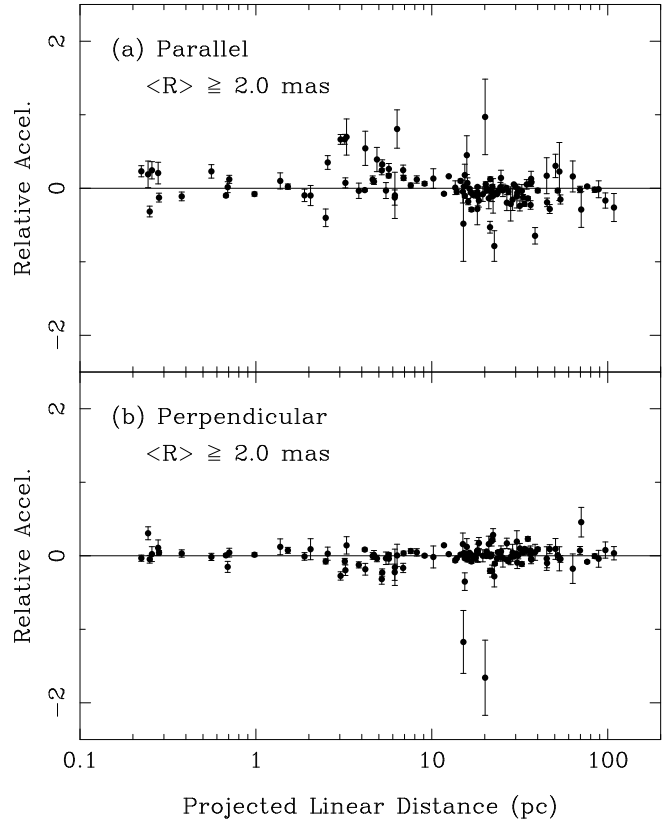}}
\caption{
\label{f:accel_v_pc_R_geq_2mas}
Relative accelerations, parallel, $\dot{\eta}_\parallel$ (panel (a)),
and perpendicular, $\dot{\eta}_\perp$ (panel (b)), to the apparent
component motion plotted against average projected linear distance of
the component from the core position in parsecs. Only components with
mean angular separations, $\langle R\rangle \geq 2.0$ mas are plotted
here. All components with an projected linear distance of $\leq 1.0$ pc
are from 0238$-$084 (NGC~1052) at $z = 0.005$. Of the 29 components
plotted between $1.0$ and $10$ parsecs, 19 are from either 3C\,120 or
BL~Lac at $z=0.033$ and $z=0.069$ respectively.
}
\end{figure}

Figure \ref{f:accel_v_R} plots relative parallel and perpendicular
accelerations versus average radial position, $\langle R\rangle$, from
the base of the jet in milli-arcseconds. There appears to be little
overall relation between average radial position and relative
acceleration, although components at less than 2 milli-arcseconds from
the base of the jet appear to have a larger spread in the magnitudes as
well as larger uncertainties for both their relative parallel and
perpendicular accelerations. K-S tests on these quantities show
significant differences between the distributions for the components at
$\langle R\rangle < 2$ milli-arcseconds compared to those at $\geq 2$
milli-arcseconds with $P \leq 0.001$ in each case. This result suggests
that components that spent a significant fraction of their time within a
few beam-widths of the base of the jet may have accelerations which are
less well determined than those at large angular radii, perhaps due to
confusion from closely spaced components in the core region and/or the
core itself. 

Figures~\ref{f:accel_v_pc} and \ref{f:accel_v_pc_R_geq_2mas} plot
relative parallel and perpendicular accelerations versus the average
projected linear distance, $d_{proj} = D_A\langle R\rangle$, for a
component, where $D_A$ is the angular size distance to the host galaxy.
Figure \ref{f:accel_v_pc_R_geq_2mas} includes only those components at
$\langle R\rangle \geq 2$ milli-arcseconds. In Figure 
\ref{f:accel_v_pc} there appears to be a decreasing spread in the
magnitudes of accelerations at larger projected distances; however, this
trend mostly disappears when the components at small angular separations
are excluded in Figure \ref{f:accel_v_pc_R_geq_2mas}. 

The relative parallel accelerations in Figure
\ref{f:accel_v_pc_R_geq_2mas} appear to show a bias towards positive
accelerations at projected distances of $\lesssim 15$ pc and a bias
toward negative accelerations at larger projected distances. These
biases also appear in Figure \ref{f:accel_v_pc} but with more scatter. 

\begin{figure}[t]
\centering
\resizebox{1.0\hsize}{!}{\includegraphics[angle=0]{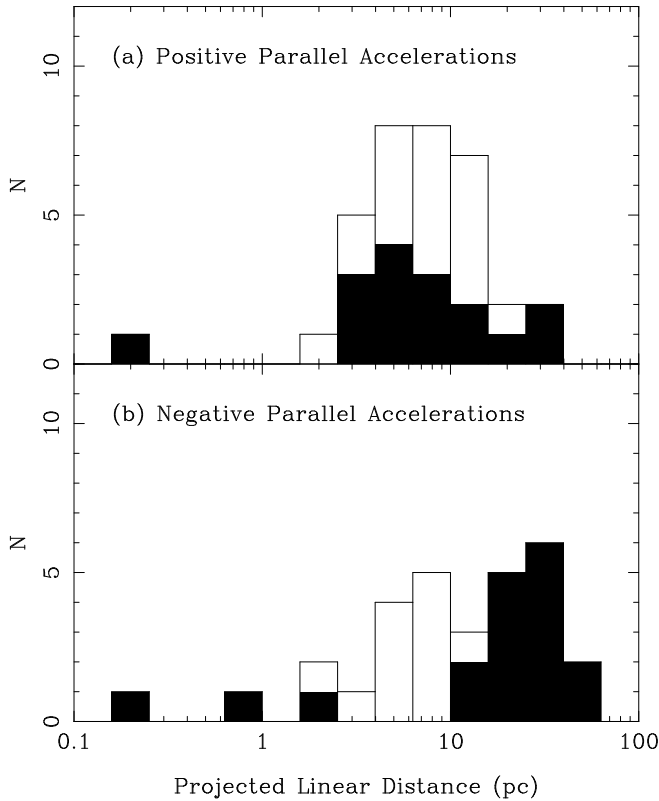}}
\caption{
\label{f:accel_v_pc_hist}
Histograms of projected linear distance for components with positive
parallel accelerations (panel (a)) and negative parallel accelerations
(panel (b)). Only components with angular accelerations significant at
the $\geq 3\sigma$ level are plotted here. Components at mean angular
separations $\langle R\rangle \geq 2.0$ milli-arcseconds are plotted as
solid bars.
}
\end{figure}

To investigate further the relationship between sign of the apparent
parallel acceleration and projected linear distance, Figure
\ref{f:accel_v_pc_hist} contains histograms of the projected linear
distances of components with positive and negative parallel
acceleration. These histograms contain only the $N=64$ components which
have $\geq 3\sigma$ parallel angular accelerations so that we may be
confident of the sign of the acceleration. A K-S test shows a
significant difference between the linear distance distributions of the
positively and negatively accelerating components, whether we consider
all components: $P = 0.013$ with mean linear distances of $9.1$ and
$17.1$ parsecs for the positively and negatively accelerating components
respectively, or we consider only components at mean angular distances
$\langle R\rangle \geq 2.0$ milli-arcseconds, where we find $P = 0.003$
with mean linear distances of $10.4$ and $24.0$ parsecs for the positive
and negatively accelerating components respectively. If we limit the
analysis to jets at larger redshifts of $z > 0.1$, we also find a
significant difference between the linear distance distributions of
positively and negatively accelerating components, with $P = 0.023$ and
mean linear distances of $10.9$ and $18.8$ parsecs for the positive and
negative components respectively. 

If we further limit the larger redshift, $z > 0.1$, sample to components
at $\langle R\rangle \geq 2.0$ milli-arcseconds, we have too few
components (7 positive accelerations, 15 negative) to reliably detect a
difference between their linear distance distributions ($P = 0.14$ with
means of $18.2$ and $28.5$ parsecs for the positive and negative
components respectively); however, if we allow $\geq 2\sigma$ parallel
angular accelerations to contribute to the test, we do find a
significant difference between the populations with $P=0.013$ that they
are drawn from the same distribution with means of $16.7$ and $28.3$
parsecs linear distance for the positive and negative components
respectively. Indeed, including the $\geq 2\sigma$ parallel angular
accelerations in all of the above tests of the linear distance
distributions increases the overall sample from $N=64$ to $N=96$
components and reduces the P-value in each test, improving our
confidence that the linear distances for the positive and negative
parallel accelerations are drawn from different distributions. 

No similar bias between positive and negative perpendicular
accelerations appears in Figures \ref{f:accel_v_pc} and
\ref{f:accel_v_pc_R_geq_2mas}, and a K-S test on the linear distance
distributions for components which have $\geq 3\sigma$ perpendicular
angular acceleration detects no difference between the positive and
negative perpendicular accelerations: $P = 0.78$ with nearly identical
means of $15.6$ parsecs.

\subsection{Connection Between Non-Radial Motion and Acceleration}
\label{s:nonrad_accel}

\begin{figure*}
\centering
\resizebox{0.96\hsize}{!}{\includegraphics[angle=-90]{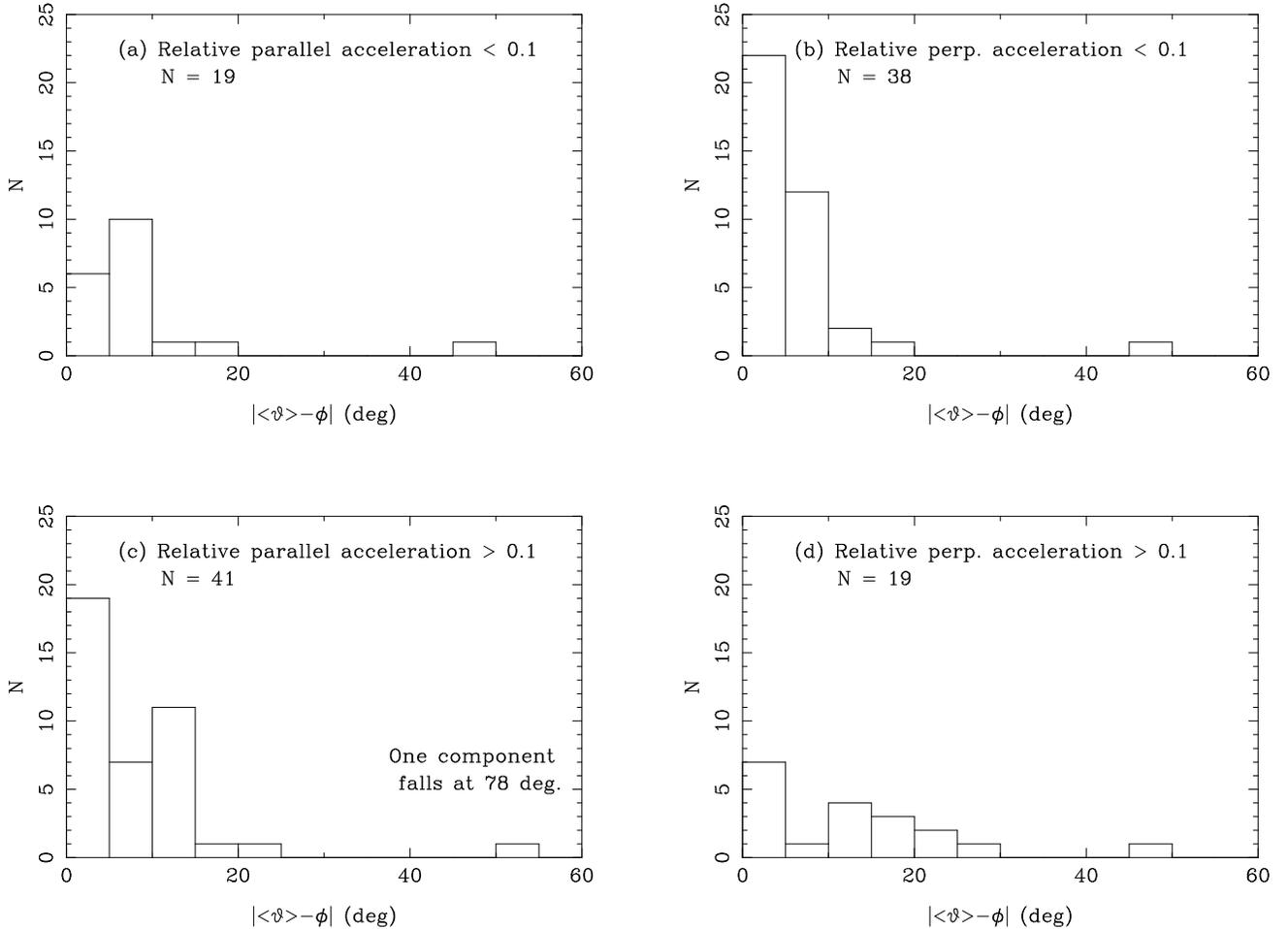}}
\caption{
\label{f:nonradial_accel_dualhist}
Histograms of velocity vector misalignments,
$|\langle\vartheta\rangle-\phi|$, for jet components with low and high
accelerations (top and bottom panels respectively) either parallel or
perpendicular to the component motion (left and right panels
respectively), as described in section \ref{s:nonrad_accel}. Note that
in panels (a) and (c), which plot parallel accelerations, components
which also have high perpendicular accelerations are not included. 
Likewise, in panels (b) and (d), which plot perpendicular 
accelerations, components which also have high parallel accelerations
are not included.
}
\end{figure*}

Figure~\ref{f:nonradial_accel_dualhist} shows the
$|\langle\vartheta\rangle-\phi|$ distributions for high acceleration and
low acceleration components for both parallel and perpendicular
accelerations. High and low acceleration components are defined as those
with relative accelerations greater or less than 0.1 at a $2\sigma$
level. To isolate the relationship of either parallel or perpendicular
acceleration to non-radial motion, components with high  parallel
acceleration are excluded from the perpendicular acceleration histograms
(panels (b) and (d)), and likewise, components with high perpendicular
acceleration are excluded from the parallel acceleration histograms
(panels (a) and (c)). A K-S test finds no significant difference in the
$|\langle\vartheta\rangle-\phi|$ distributions for parallel
accelerations: $P = 0.12$ with mean values of $7.8$ and $9.6$ degrees
for the low and high parallel accelerating components respectively.
However, a K-S test does detect a highly significant difference between
the distributions for perpendicular accelerations: $P=0.004$ with mean
values of $5.1$ and $12.6$ degrees for the low and high perpendicular
accelerations respectively. These results indicate that strong
perpendicular accelerations are linked with higher degrees of non-radial
motion.

\begin{figure}[t]
\centering
\resizebox{1.0\hsize}{!}{\includegraphics[angle=-90]{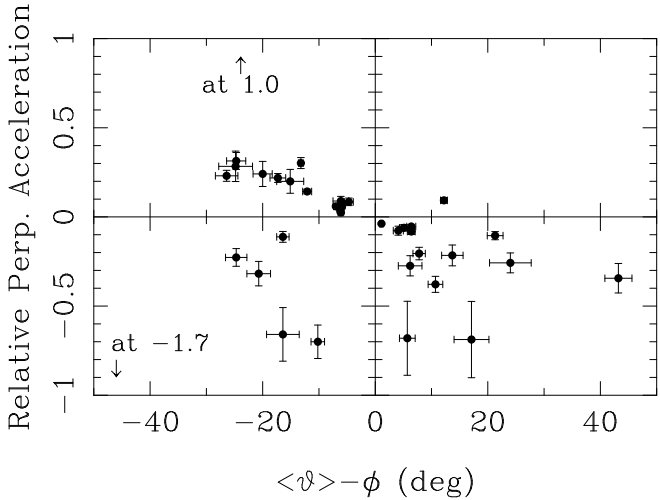}}
\caption{
\label{f:accel_v_offset}
Plot of relative perpendicular
acceleration versus $\langle\vartheta\rangle-\phi$ for all jet
components with both significant non-radial motion and perpendicular
acceleration in Table~\ref{t:comps}. Of the 37 components plotted here,
30 have opposite signs for their relative perpendicular accelerations
and $\langle\vartheta\rangle-\phi$, indicating that the accelerations
are in the correct direction to have produced the observed non-radial
motion. Note that two points, indicated by arrows, have relative
accelerations that fall above and below the plot at $+1.0$ and $-1.7$
respectively.
}
\end{figure}

Figure~\ref{f:accel_v_offset}, which plots the relative perpendicular
acceleration, $\dot{\eta}_\perp$, versus the proper motion angle
misalignment, $\langle\vartheta\rangle-\phi$, shows this relationship
further. Note that here we have not taken the absolute value of the
misalignment angle so that we can compare the sign of the non-radial
motion with the sign of the perpendicular acceleration.
Figure~\ref{f:accel_v_offset} includes all 37 components from
Table~\ref{t:comps} which have both a significant non-radial motion {\em
and} a significant perpendicular acceleration. We find that 30 of the 37
components have a perpendicular acceleration in the correct direction to
produce the observed sign of the proper motion offset; i.e., with our
sign definitions, a positive perpendicular acceleration in an originally
radially moving component will eventually produce a negative
misalignment. The probability of this agreement occurring by pure chance
is extremely small ($P=0.00015$). Figure~\ref{f:accel_v_offset} also
shows a possible trend to larger accelerations for larger misalignments,
although there are several exceptions.

\section{Discussion}
\label{s:discuss}

In the previous section we presented the results of our analysis of
accelerations measured from 203 jet components from the MOJAVE sample.
We reported and analyzed acceleration measurements both parallel and
perpendicular to the apparent velocity. Parallel accelerations
characterize changes in apparent speed of a component while
perpendicular accelerations characterize changes in direction on the
sky. We found significant angular accelerations parallel to the
component velocities in about one-third of the sample, and about
one-fifth had significant perpendicular accelerations. To facilitate
comparison of accelerations between different components and jets, we
constructed relative accelerations, $\dot{\eta}_\parallel$ and
$\dot{\eta}_\perp$, where we divide each acceleration by the apparent
velocity of the component. As noted in \S{2}, relative accelerations are
also convenient for investigating rates of change of the intrinsic
quantities that may lead to the observed accelerations. 

\subsection{Changes in Speed}

Parallel accelerations, representing changes in apparent speed of a
component or pattern, can be due either to intrinsic changes in the
Lorentz factor, $\Gamma$, or angle to the line of sight, $\theta$, see
\S{\ref{s:parallel}}. Distinguishing between these possibilities in any
individual case is difficult; however, we can begin by making the simple
assumption that all observed accelerations are due only to changes in
angle: $\theta$ and/or $\phi$. Under this assumption, we showed in
\S{\ref{s:accel_ratio}} that, averaged across many components and jets,
we expect that, on average, the observed parallel accelerations would be
about 60\% of the magnitude of the observed perpendicular accelerations.
However, this is not what we observed. We found that the distribution of
parallel accelerations was much broader than the distribution of
perpendicular accelerations with the mean parallel acceleration larger
than the mean perpendicular acceleration by about a factor of $1.6$. 

Therefore, we infer that more than half of the magnitudes of the
observed parallel accelerations can be attributed to intrinsic changes
in the Lorentz factor, $\Gamma$, of the component or pattern. We
emphasize that this is only an average statement, and that for any given
component, we cannot be certain how much, if any, of the observed
parallel acceleration is due to Lorentz factor versus intrinsic angle
changes. We note, however, that for components moving at the critical
angle for maximal superluminal motion, $\cos\theta=\beta$, equation
\ref{e:dbeta_obs1} shows that any observed parallel acceleration must be
due to changes in the Lorentz factor alone. 

Given these results and the measurement of significant parallel
accelerations in roughly a third of our sample, we conclude that changes
in intrinsic speed are common. Whether these changes in component or
pattern motion reflect underlying changes in the flow speed of the jet
is not clear, and we discuss this issue further below.

Regarding the magnitudes of the parallel accelerations, in
\S{\ref{s:accel_results}} we characterized a relative acceleration of $>
0.1$ as high, corresponding to a more than 10\% change in the component
velocity during one year in our frame. We found that roughly a quarter
($N=51$) of our components had relative parallel accelerations at least
$2\sigma$ greater than $0.1$, and a few components have much larger
relative parallel accelerations: nine are greater than $0.5$ at the
$2\sigma$ level. If a jet component is moving with $\Gamma = \delta =
10$ and a relative parallel acceleration of $0.1$, this corresponds to
an intrinsic rate of change characterized by $\dot{\Gamma}/\Gamma =
10^{-3}$ per year in the frame of the AGN host galaxy (see
\S{\ref{s:parallel}}).

\subsubsection{Relation with Linear Distance}

As described in detail in \S{\ref{s:accel_results}, we found that
components with positive parallel accelerations tended to appear at
shorter linear distances from the jet core while components with
negative parallel accelerations tended to appear at longer linear
distances with means of approximately $10$ and $20$ parsecs
respectively. This trend can be seen most clearly in the histogram given
in Figure \ref{f:accel_v_pc_hist} where the difference is strongest for
those components at mean angular separations $\langle R\rangle \geq 2.0$
milli-arcseconds. We found that these differences persist even if we
consider only jets at redshifts $z > 0.1$, eliminating a few nearby jets
with several components at shorter projected distances.

While this trend is robust, positively accelerating components appear at
shorter projected distances on average than negatively accelerating
components, we note that there are many exceptions to the trend and no
single source shows this clear pattern in several components all by
itself. More typical for jets with multiple components is that some of
the components show acceleration (either positive, negative, or both)
and others do not. Given the arguments in the previous section, we
expect that a significant fraction of these parallel accelerations
reflect real changes in the Lorentz factor of the components or
patterns; however, we do not know the extent to which they reflect
changes in the underlying flow. If they do, then we may be seeing an
overall trend for positive acceleration of the flow at shorter projected
distances, $\lesssim 15$ parsecs, and negative acceleration of the flow
at longer projected distances, $\gtrsim 15$ parsecs. 

Quasar jets are observed to be at least mildly relativistic on scales of
hundreds of kiloparsecs with beaming models of the observed jet and 
counterjet radio emission giving $\Gamma_{kpc} \simeq 1.3$ with values
$\gtrsim 3$ being inconsistent with the data, suggesting at least some
deceleration between parsec and kiloparsec scales \citep{WA97,H99,MH09};
however, the interpretation of X-ray emission observed on kiloparsec
scales as inverse-Compton scattering from the cosmic microwave
background seems to require $\Gamma_{kpc} \simeq 3-15$
\citep[e.g][]{S04}.

If some jets begin decelerating on decaparsec scales, the rate of
deceleration cannot remain at the relatively high values seen here. A
jet component moving with $\Gamma=\delta=10$ and an observed
deceleration $\dot{\eta}_\parallel = -0.1$ will be experiencing a change
in Lorentz factor characterized by $\dot{\Gamma}/\Gamma = -10^{-3}$ per
year in the rest frame of the host galaxy. If this rate remains
constant, the component will decelerate to $\Gamma < 1.3$ in just 2000
years, long before it can travel hundreds of kiloparsecs.

\subsubsection{Pattern Speed versus Flow Speed}

In Paper VI we showed that the range of component speeds within an
individual jet was much smaller than the range of component speeds
across our sample as a whole, indicating that individual jets do have a
characteristic flow speed. However, in many cases, the range of speeds
observed within individual jets was still reasonably large. In the
sample of 203 component motions analyzed here, a range of speeds
spanning a factor of two in an individual jet is not uncommon, and, in a
few cases, there are larger differences. Taken together these result
suggests that (1) pattern motion at a different speed than the
underlying flow is a common feature of the component motion we are
studying, and that (2) the observed pattern motions do carry information
about the underlying flow or we would see a much wider spread of
apparent speed in individual jets. A key question is the extent to which
the changes observed in component or pattern motions are reflective of
changes in the underlying flow.

We noted above that the trend for components with increasing speed to be
at smaller projected distances than components with decreasing speed was
robust; however, there were many exceptions and individual jets did not
show clear evidence of this pattern by themselves. These exceptions are
consistent with some contribution from both random changes in pattern
motion and changes in direction to the line of sight; however, the
robustness of the overall trend suggests a common physical link.  That
link may be to the underlying flow, as suggested above, reflecting a
tendency for positive acceleration of the flow speed at shorter
projected distances, $\lesssim 15$ parsecs, and negative acceleration of
the flow speed at longer projected distances, $\gtrsim 15$ parsecs. 
Alternatively, there may be common hydro-dynamical processes for shock
propagation, such as the expectation that a leading shock will slowly
decrease in apparent speed \citep{MG85}, which also play a role in
producing this trend.

\subsection{Changes in Direction}

We characterized the 'non-radial' motion for each component by its
misalignment angle, the difference between the component's structural
position angle, $\langle\vartheta\rangle$, and its vector motion
position angle, $\phi$. Components with a misalignment angle of zero
appear to be moving radially or ballistically, while those with 
significant misalignments must have changed their trajectory at some
point. We found that significant non-radial motions were common,
appearing in about half of our sample, and that large non-radial motions
with misalignment angles robustly greater than $10^\circ$ occur in about
one-fifth of our sample. In general, the non-radial motions we observed
were in the direction of the downstream jet, confirming the results of
\citet{KL04} that while jet components may have different position
angles, they ultimately tend to follow a pre-established flow direction
rather than continue on ballistic trajectories. The degree to which this
tendency results from jet collimation on parsec scales or simple bending
of an already well collimated jet is unclear.

While the misalignment angles described above give a indication of
whether a component has changed its trajectory in the past, they are not
a direct measure of how much the component motion may have changed
during our observations. However, the observed perpendicular
accelerations for each component provide just such a direct measure of
changes in direction of apparent component motion. 

In \S\ref{s:nonrad_accel} we showed there was a close connection
between the observed perpendicular accelerations and the observed
misalignment angles that characterize non-radial motion. We found that
components with high perpendicular acceleration had a broader
distribution of misalignment angles. We also examined components that
had both a significant misalignment and a significant perpendicular
acceleration, and we found the observed perpendicular acceleration had a
strong tendency to be in the correct direction to have caused the
observed misalignment angle. This is precisely what we would expect for
a scenario in which components must have experienced some acceleration
in the past to become non-radial.

\subsection{Changes in Acceleration}
\label{s:limits}

As discussed briefly in \S{\ref{s:results}} and in more detail in Paper
VI, our proper motion model assumes a single, constant acceleration term
for each of the $x$ and $y$ position coordinates in the sky. We then
resolve these into accelerations parallel, $\dot{\mu}_\parallel$, and
perpendicular, $\dot{\mu}_\perp$ to the observed velocity. The
accelerations are fit relative to the middle epoch of a given
component's observations, and therefore represent the average
accelerations experienced by the component throughout the period of
observation. 

With this approach we are not sensitive to changes in acceleration, or
``jerks'', that might be experienced by a component during our
observations. A key example is the sudden change in speed and direction
reported by \citet{H03} in a strong component in 3C\,279.  During 1998,
this component increased its apparent speed by more than 50\% and
changed its direction on the sky by $\sim 25^\circ$. Just prior to the
change in direction, the component had a sharp rise in flux density and
decrease in transverse size, and it had a steady decrease in flux
density after the change \citep{H03}. \citet{ZT01} also reported a
change in polarization of the component associated with this change in
direction. In the results reported here, this large, sudden change is
just one part of a much longer trajectory for this component
($1253-055$, component 1 in Table \ref{t:comps}) spanning 89 VLBA epochs
at $15$ GHz since 1994. If we examine the trajectory from 1999 onward,
after the sudden change, the remaining 77 epochs show significant
parallel and perpendicular accelerations: $\dot{\mu}_\parallel =
-5.6\pm1.7$ and $\dot{\mu}_\perp = +13.0\pm1.6$ micro-arcsec/yr/yr,
which differ in sign and magnitude from the average values for the
entire trajectory reported in Table \ref{t:comps}. 

Changes in accelerations of individual components might result not only
from discrete, impulsive events, like that seen in 3C\,279, but also
from more gradual changes, perhaps as a component moves from a region of
increasing flow speed to a region of decreasing flow speed. Our $89$
epochs, spanning 13 years, on component 1 in 3C\,279 represents an
extreme example in temporal coverage of a single moving component. 
Improved time baselines on many other components may permit a larger
study for changes in acceleration in individual components in the
future. This issue might also be addressed with a detailed study
examining changes in apparent velocity with other component properties
such as flux density, polarization, and size, but that is beyond the
scope of this work.

\section{Summary and Conclusions}

We have analyzed acceleration measurements for 203 jet components from
the proper motions reported for the MOJAVE program in Paper VI. We have
examined accelerations both parallel and perpendicular to the component
velocity, representing changes in apparent speed and direction
respectively. Our main results are as follows:

(i) We find significant parallel accelerations in roughly one third of
our sample and significant perpendicular accelerations in about one
fifth of our sample.

(ii) To compare accelerations between components moving with different
apparent speeds, we define relative accelerations by taking the ratio
$\dot{\beta}_{obs}/\beta_{obs}$. Defining high relative accelerations as
those with magnitude $> 0.1$, corresponding to a 10\% change in apparent
velocity per year in our frame, we find that about one quarter of the
components in our sample show high parallel accelerations and about one
seventh have high perpendicular accelerations.

(iii) Changes in the apparent speed of components are due to changes in
their Lorentz factor, as much or more so, than due to changes in their
angle to the line of sight. This is evidenced by the fact that parallel
accelerations have larger magnitudes on average than perpendicular
accelerations. We estimate that more than half of the observed parallel
accelerations are due to intrinsic changes in speed of the components or
pattern.

(iv) On average, components with positive parallel accelerations,
representing increasing apparent speed, are found closer to the jet base
than components with negative parallel accelerations. While numerous
exceptions to this trend are observed, perhaps due to the influence of
pattern motion or directional changes, the overall trend itself is
robust, suggesting a physical link between the observed pattern motions
and the underlying flow. This trend may reflect a tendency for the jet
flow to increase in speed on shorter length scales ($\lesssim 15$ pc
projected) and decrease in speed at longer distances; however common
hydro-dynamical processes for shock propagation may also play a role in
producing the observed trend. 

(v) Significant non-radial motions, characterized by a misalignment
between a component's structural position angle and its velocity
direction, appear in half of our sample with about one fifth showing
significant misalignments greater than $10^\circ$. These misalignments
have a strong tendency to be in the direction to better align the
component motion with the downstream flow, confirming the results of
\citet{KL04} that new jet components tend to follow a pre-established
flow direction rather than continue on ballistic trajectories from the
jet core. 

(vi) Perpendicular accelerations appear to be closely linked with
misalignment angles and are usually in the correct direction to have
caused the observed misalignment.

Overall, the high incidence of accelerations and non-radial motions in
our observations indicates that jet flows are still becoming organized
on parsec to decaparsec scales. However, with a few exceptions, the
observed changes, magnified by projection and apparent time compression,
suggest only modest rates of change in intrinsic parameters on these
length scales.

\acknowledgements
The authors wish to acknowledge the other members of the MOJAVE team.
DCH has been supported under NSF grant AST-0707693. MLL has been
supported under NSF grants AST-0406923 \& AST-0807860,
NASA-\textit{Fermi} grant NNX08AV67G and a grant from the Purdue
Research Foundation. TS has been supported in part by the Academy of
Finland grant 120516. MK has been supported in part by an appointment to
the NASA Postdoctoral Program at the Goddard Space Flight Center,
administered by Oak Ridge Associated Universities through a contract
with NASA. Part of this work was done by YYK and TS during their
Alexander von Humboldt fellowships at the MPIfR. YYK is partly supported
by the Russian Foundation for Basic Research (project 08-02-00545). This
work has made use of data obtained from the National Radio Astronomy
Observatory's Very Long Baseline Array and its public archive. The
National Radio Astronomy Observatory is a facility of the National
Science Foundation operated under cooperative agreement by Associated
Universities, Inc. This research has made use of NASA's Astrophysics
Data System, and the NASA/IPAC Extragalactic Database (NED). The latter
is operated by the Jet Propulsion Laboratory, California Institute of
Technology, under contract with the National Aeronautics and Space
Administration. 

\appendix

\section{Changes in Apparent Motion}

This appendix complements section 2 by presenting more detailed
background and derivations of the relationships describing changes in
apparent speed of parsec-scale AGN jet features observed in VLBI
monitoring. This work builds upon the relationships laid out in
\citet{BK79} and sources referenced therein. All of the relationships
derived here apply strictly to the moving patterns that we see as
features or ``components'' propagating in a jet. Intrinsically, a moving
pattern has speed = $\beta c$ at angle $\theta$ to the line of sight. In
principle, the flow of the jet may have a different speed, $\beta_f c$,
and angle, $\theta_f$; however, beyond differences due to Doppler
boosting, these values do not affect the apparent motions or changes in
motion of the patterns. Here we concern ourselves with pattern motion as
the key observable. Note that for economy of notation, we have not
applied a subscript to indicate the pattern quantities ($\beta$,
$\Gamma$, $\theta$,etc...) as additional subscripts are necessary for
describing other aspects of the motion.

\subsection{Basic Time Derivatives}
\label{s:time}

We are interested in relating the rate of change of the observed angular
speed, $\mu$, to the intrinsic rates of change of the pattern speed (or
equivalently, Lorentz factor) and angle to the line of sight. To make
these comparisons, we first define the time frames involved and 
derivatives of the corresponding quantities.

We consider three time frames as follows: 
\begin{itemize}
\item 
Time, $t$, is taken to be the time in the host galaxy of the jet. This
time does not include apparent time compression due to motion along the
line of sight. It is the appropriate time for defining derivatives of
the intrinsic quantities like Lorentz factor and angle to the line of
sight. 
\item 
We define the time $t_{obs}$ to be the time frame in which events appear
to unfold as seen along the direction of the observer in the host
galaxy. This time does include apparent time compression due to motion
along the line of sight, where $dt_{obs} = (1-\beta\cos\theta)dt$.
\item
We also define $t'_{obs}$ to be the time in the frame of the observer in
our galaxy, so $dt'_{obs} = (1+z)dt_{obs}$.
\end{itemize}

With these time frames defined, we define the ``dot'' notation for
derivatives to mean a time derivative with respect to the appropriate
time frame for the quantity in question. Thus we have the following
definitions for time derivatives of the basic quantities:
\begin{equation}
\dot{\beta} = \frac{d\beta}{dt}
\end{equation}
\begin{equation}
\dot{\Gamma} = \frac{d\Gamma}{dt} = \dot{\beta} \beta \Gamma^3  
\end{equation}
\begin{equation}
\dot{\beta}_{obs} = \frac{d\beta_{obs}}{dt_{obs}}
\end{equation}
\begin{equation}
\dot{\mu} = \frac{d\mu}{dt'_{obs}} = \frac{d\mu}{dt_{obs}} (1+z)^{-1}
\end{equation}

\subsection{Vector Acceleration}

\begin{figure}
\centering
\resizebox{0.9\hsize}{!}{\includegraphics[angle=0]{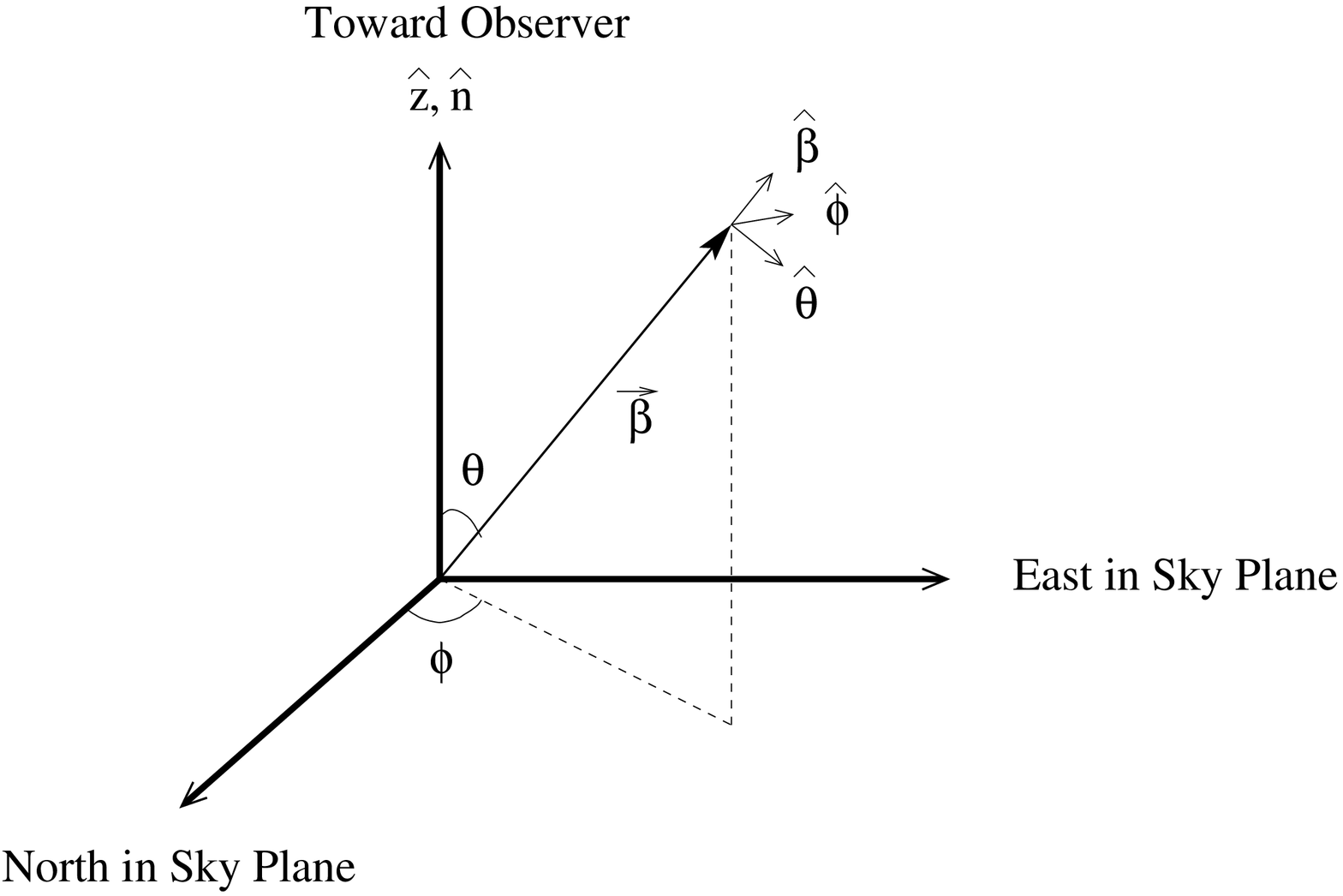}}
\figurenum{A1}
\caption{
\label{f:beta_geom}
Three dimensional geometry of the velocity vector, $\vec{\beta}$, which
is taken in the text to be a vector in a standard spherical-polar
coordinate system, as illustrated. The unit vectors
$\hat{\beta},\hat{\phi},\hat{\theta}$ are mutually orthogonal.
}
\end{figure}

To allow the possibility of changes in both the speed and direction of a
jet feature's motion we consider the full vector acceleration, equation
(3) from \citet{BK79}:
\begin{equation}
\label{e:vect_accel}
\frac{d\vec{\beta}_{obs}}{dt_{obs}} = (1-\vec{\beta}\cdot\hat{n})^{-3}\left[(1-\vec{\beta}\cdot\hat{n})
d\vec{\beta}/dt+(d\vec{\beta}/dt\cdot\hat{n})(\vec{\beta}-\hat{n})\right]
\end{equation}
where $\hat{n}$ is the direction toward the observer, so
$(1-\vec{\beta}\cdot\hat{n})$ is the familiar factor,
$(1-\beta\cos\theta)$. In the analysis that follows, we take 
$\vec{\beta} = \beta\hat{\beta}$ to be a vector in a spherical-polar
coordinate system with the z-axis oriented towards the observer, such
that $\hat{n}=\hat{z}=\cos\theta\hat{\beta}-\sin\theta\hat{\theta}$, see
Figure \ref{f:beta_geom}. The unit vector, $\hat{\theta}$, defines the
direction of increasing angle to the line of sight, and the angle $\phi$
is the polar angle in the plane of the sky. In these coordinates,
$d\vec{\beta}/dt =
\dot{\beta}\hat{\beta}+\beta\dot{\phi}\sin\theta\hat{\phi}+\beta\dot{\theta}\hat{\theta}$,
where we note that $\hat{\phi}$ is perpendicular to both $\hat{\beta}$
and $\hat{\theta}$ and is therefore always perpendicular to the observed
motion in the plane of the sky. 

To find the observed parallel acceleration, along the direction of
observed motion in the plane of the sky, we take the dot product of eqn.
\ref{e:vect_accel} with the unit vector along $\vec{\beta}_{obs}$,
$\hat{\beta}_{obs} = \sin\theta\hat{\beta}+\cos\theta\hat{\theta}$. 
\begin{equation}
\frac{d\beta_{\parallel obs}}{dt_{obs}} = \frac{d\vec{\beta}_{obs}}{dt_{obs}}\cdot\hat{\beta}_{obs}
= \frac{\dot{\beta}\sin\theta+\beta\dot{\theta}(\cos\theta-\beta)}{(1-\beta\cos\theta)^3}
\end{equation}
As expected, this expression matches what we would get by simply
differentiating the familiar apparent speed expression given in equation
\ref{e:beta_obs} and correcting for the appropriate time frames as
described in the previous section. 

To find the observed perpendicular acceleration, transverse to the
direction of observed motion in the plane of the sky, we simply take the
dot product of eqn. \ref{e:vect_accel} with $\hat{\phi}$:
\begin{equation}
\frac{d\beta_{\perp obs}}{dt_{obs}} =\frac{d\vec{\beta}_{obs}}{dt_{obs}}\cdot\hat{\phi} 
 = \frac{\beta\dot{\phi}\sin\theta}{(1-\beta\cos\theta)^2},
\end{equation}
so the observed perpendicular acceleration is only related to changes
in the angle $\phi$, as we might expect.


\clearpage
\LongTables
\tabletypesize{\scriptsize} 
\begin{deluxetable*}{llrrrrrrrrrrrr} 
\tablecolumns{14} 
\tabletypesize{\scriptsize} 
\tablewidth{0pt}  
\tablecaption{
\label{t:comps} 
Components for Acceleration and Non-Radial Motion Analysis}  
\tablehead{
\colhead{} &   \colhead {} & \colhead{} & \colhead{$\langle R\rangle$} &
\colhead{$\langle\vartheta\rangle$} & \colhead {$d_{proj}$} &   \colhead {$\mu$} &  \colhead{$\beta_{obs}$} &
\colhead{$\phi$} & \colhead{$|\langle\vartheta\rangle-\phi|$} &
\colhead{$\dot{\mu}_\parallel$} & \colhead {$\dot{\mu}_\perp$} &   \colhead {} & \colhead{} \\
\colhead{Source} &   \colhead {I.D.} & \colhead{N} & \colhead{(mas)} &
\colhead{(deg.)} & \colhead {(pc)} &   \colhead {($\mu$as yr$^{-1}$)} & & \colhead{(deg.)} & \colhead{(deg.)} &
\colhead{($\mu$as yr$^{-2}$)} & \colhead {($\mu$as yr$^{-2}$)} &   \colhead {$\dot{\eta}_\parallel$} & \colhead{$\dot{\eta}_\perp$} \\
\colhead{(1)} & \colhead{(2)} & \colhead{(3)} & \colhead{(4)} &  \colhead{(5)} &
\colhead{(6)} & \colhead{(7)} & \colhead{(8)} & \colhead{(9)} &  \colhead{(10)} &
\colhead{(11)} & \colhead{(12)} & \colhead{(13)} & \colhead{(14)} 
}
\startdata 
\scriptsize
0016+731 & $1$ & $12$ & $1.10$ & $137.3$ & $9.40$ & $87\pm5$ & $6.7\pm0.4$ &  $152.1\pm2.6$ & $14.7\pm2.8$ & $4\pm3$ & $5\pm2$ & $0.14\pm0.09$ & $0.16\pm0.07$ \\
0059+581 & $2$ & $13$ & $1.46$ & $-119.2$ & $10.07$ & $197\pm14$ & $7.3\pm0.5$ &  $-106.7\pm4.3$ & $12.5\pm4.5$ & $49\pm13$ & $-2\pm15$ & $0.41\pm0.11$ & $-0.01\pm0.12$ \\
         & $3$ & $10$ & $0.81$ & $-125.8$ & $5.58$ & $283\pm20$ & $10.5\pm0.7$ &  $-111.3\pm3.8$ & $14.5\pm4.0$ & $-20\pm45$ & $-14\pm43$ & $-0.11\pm0.26$ & $-0.08\pm0.25$ \\
         & $4$ & $10$ & $0.59$ & $-163.8$ & $4.07$ & $162\pm13$ & $6.0\pm0.5$ &  $-129.3\pm4.8$ & $34.4\pm5.0$ & $-35\pm29$ & $10\pm30$ & $-0.36\pm0.29$ & $0.10\pm0.31$ \\
0202+319 & $1$ & $10$ & $7.42$ & $-5.9$ & $63.30$ & $121\pm11$ & $8.2\pm0.8$ &  $-38.7\pm5.0$ & $32.8\pm5.0$ & $8\pm10$ & $-9\pm10$ & $0.16\pm0.21$ & $-0.18\pm0.20$ \\
0212+735 & $1$ & $14$ & $0.84$ & $115.8$ & $6.95$ & $84\pm3$ & $7.5\pm0.3$ &  $102.0\pm1.8$ & $13.7\pm1.9$ & $-6\pm2$ & $-5\pm1$ & $-0.23\pm0.06$ & $-0.22\pm0.06$ \\
0215+015 & $2$ & $12$ & $2.12$ & $107.2$ & $18.14$ & $451\pm28$ & $33.8\pm2.1$ &  $104.5\pm2.2$ & $2.7\pm2.4$ & $-47\pm36$ & $11\pm24$ & $-0.28\pm0.21$ & $0.06\pm0.14$ \\
0234+285 & $3$ & $15$ & $3.89$ & $-11.8$ & $32.50$ & $199\pm11$ & $11.9\pm0.7$ &  $4.7\pm1.1$ & $16.4\pm1.1$ & $-3\pm8$ & $-10\pm3$ & $-0.04\pm0.09$ & $-0.11\pm0.03$ \\
0238$-$084 & $3$ & $24$ & $9.46$ & $68.8$ & $0.98$ & $594\pm23$ & $0.2\pm0.0$ &  $67.1\pm2.0$ & $1.7\pm2.1$ & $-47\pm16$ & $8\pm16$ & $-0.08\pm0.03$ & $0.01\pm0.03$ \\
\,(NGC 1052) & $4$ & $14$ & $6.80$ & $72.3$ & $0.71$ & $873\pm37$ & $0.3\pm0.0$ &  $74.7\pm2.2$ & $2.3\pm2.2$ & $105\pm47$ & $38\pm51$ & $0.12\pm0.05$ & $0.04\pm0.06$ \\
         & $5$ & $12$ & $5.38$ & $69.6$ & $0.56$ & $746\pm48$ & $0.3\pm0.0$ &  $69.9\pm2.0$ & $0.3\pm2.0$ & $170\pm67$ & $-11\pm36$ & $0.23\pm0.09$ & $-0.02\pm0.05$ \\
         & $10$ & $10$ & $2.35$ & $68.6$ & $0.24$ & $986\pm50$ & $0.3\pm0.0$ &  $72.2\pm1.4$ & $3.7\pm1.5$ & $185\pm178$ & $300\pm85$ & $0.19\pm0.18$ & $0.31\pm0.09$ \\
         & $11$ & $13$ & $2.71$ & $67.7$ & $0.28$ & $804\pm34$ & $0.3\pm0.0$ &  $71.0\pm1.4$ & $3.3\pm1.5$ & $-102\pm48$ & $36\pm22$ & $-0.13\pm0.06$ & $0.04\pm0.03$ \\
         & $12$ & $12$ & $2.40$ & $66.5$ & $0.25$ & $696\pm26$ & $0.2\pm0.0$ &  $67.9\pm1.6$ & $1.3\pm1.6$ & $-219\pm50$ & $-33\pm36$ & $-0.32\pm0.07$ & $-0.05\pm0.05$ \\
         & $14$ & $10$ & $1.57$ & $66.2$ & $0.16$ & $595\pm36$ & $0.2\pm0.0$ &  $67.3\pm1.7$ & $1.2\pm1.9$ & $25\pm99$ & $2\pm48$ & $0.04\pm0.17$ & $0.00\pm0.08$ \\
         & $17$ & $10$ & $3.65$ & $68.3$ & $0.38$ & $1033\pm31$ & $0.4\pm0.0$ &  $71.3\pm1.4$ & $3.0\pm1.5$ & $-114\pm61$ & $32\pm52$ & $-0.11\pm0.06$ & $0.03\pm0.05$ \\
         & $26$ & $10$ & $2.68$ & $-114.7$ & $0.28$ & $751\pm56$ & $0.3\pm0.0$ &  $-111.7\pm2.8$ & $3.0\pm3.0$ & $154\pm110$ & $82\pm79$ & $0.21\pm0.15$ & $0.11\pm0.11$ \\
         & $30$ & $13$ & $2.15$ & $-111.4$ & $0.22$ & $807\pm42$ & $0.3\pm0.0$ &  $-115.2\pm1.7$ & $3.8\pm1.9$ & $185\pm60$ & $-26\pm34$ & $0.23\pm0.08$ & $-0.03\pm0.04$ \\
         & $31$ & $10$ & $2.47$ & $-114.6$ & $0.26$ & $819\pm57$ & $0.3\pm0.0$ &  $-121.1\pm3.3$ & $6.5\pm3.5$ & $199\pm95$ & $19\pm82$ & $0.24\pm0.12$ & $0.02\pm0.10$ \\
         & $33$ & $21$ & $6.51$ & $-111.9$ & $0.68$ & $773\pm24$ & $0.3\pm0.0$ &  $-112.6\pm1.7$ & $0.7\pm1.8$ & $-77\pm16$ & $5\pm15$ & $-0.10\pm0.02$ & $0.01\pm0.02$ \\
         & $34$ & $13$ & $6.66$ & $-112.8$ & $0.69$ & $1012\pm46$ & $0.3\pm0.0$ &  $-117.5\pm2.8$ & $4.7\pm2.9$ & $15\pm75$ & $-152\pm75$ & $0.01\pm0.07$ & $-0.15\pm0.08$ \\
         & $35$ & $18$ & $14.65$ & $-114.7$ & $1.52$ & $751\pm42$ & $0.3\pm0.0$ &  $-109.4\pm3.7$ & $5.3\pm3.7$ & $16\pm27$ & $55\pm30$ & $0.02\pm0.04$ & $0.07\pm0.04$ \\
0316+413 & $1$ & $22$ & $13.74$ & $178.3$ & $4.87$ & $266\pm50$ & $0.3\pm0.1$ &  $173.2\pm3.0$ & $5.1\pm3.1$ & $102\pm38$ & $-10\pm11$ & $0.39\pm0.16$ & $-0.04\pm0.04$ \\
\,(3C\,84) &&&&&&&&&&&&& \\
0333+321 & $6$ & $17$ & $3.56$ & $123.7$ & $29.93$ & $206\pm3$ & $12.7\pm0.2$ &  $125.0\pm0.9$ & $1.4\pm1.0$ & $2\pm2$ & $3\pm2$ & $0.02\pm0.03$ & $0.03\pm0.03$ \\
\,(NRAO 140) & $7$ & $17$ & $5.57$ & $127.9$ & $46.82$ & $188\pm7$ & $11.6\pm0.4$ &  $136.9\pm2.3$ & $8.9\pm2.3$ & $-24\pm5$ & $8\pm5$ & $-0.28\pm0.06$ & $0.09\pm0.06$ \\
0336-019 & $4$ & $10$ & $1.39$ & $49.9$ & $10.68$ & $253\pm16$ & $11.7\pm0.7$ &  $56.9\pm4.0$ & $7.0\pm4.4$ & $8\pm17$ & $-11\pm19$ & $0.06\pm0.13$ & $-0.08\pm0.14$ \\
0415+379 & $7$ & $12$ & $9.59$ & $64.7$ & $9.10$ & $1677\pm41$ & $5.5\pm0.1$ &  $64.5\pm1.0$ & $0.2\pm1.1$ & $101\pm42$ & $2\pm31$ & $0.06\pm0.03$ & $0.00\pm0.02$ \\
\,(3C\,111)& $9$ & $11$ & $7.29$ & $65.5$ & $6.91$ & $1269\pm31$ & $4.2\pm0.1$ &  $66.3\pm1.3$ & $0.8\pm1.4$ & $167\pm38$ & $41\pm35$ & $0.14\pm0.03$ & $0.03\pm0.03$ \\
0430+052 & $1$ & $14$ & $9.74$ & $-108.0$ & $6.35$ & $2187\pm93$ & $4.9\pm0.2$ &  $-101.8\pm1.4$ & $6.2\pm1.4$ & $1705\pm547$ & $11\pm323$ & $0.81\pm0.26$ & $0.01\pm0.15$ \\
\,(3C\,120)& $2$ & $22$ & $6.38$ & $-106.8$ & $4.16$ & $1730\pm21$ & $3.9\pm0.1$ &  $-102.1\pm0.7$ & $4.7\pm0.8$ & $-43\pm33$ & $142\pm35$ & $-0.03\pm0.02$ & $0.08\pm0.02$ \\
         & $3$ & $15$ & $7.20$ & $-116.1$ & $4.69$ & $2307\pm27$ & $5.1\pm0.1$ &  $-113.2\pm0.9$ & $2.9\pm0.9$ & $202\pm86$ & $27\pm133$ & $0.09\pm0.04$ & $0.01\pm0.06$ \\
         & $4$ & $13$ & $5.03$ & $-116.5$ & $3.28$ & $1940\pm91$ & $4.3\pm0.2$ &  $-108.9\pm1.2$ & $7.5\pm1.2$ & $1309\pm458$ & $267\pm221$ & $0.70\pm0.25$ & $0.14\pm0.12$ \\
         & $5$ & $15$ & $3.93$ & $-123.8$ & $2.56$ & $1772\pm45$ & $3.9\pm0.1$ &  $-115.5\pm1.4$ & $8.2\pm1.4$ & $601\pm160$ & $50\pm153$ & $0.35\pm0.09$ & $0.03\pm0.09$ \\
         & $6$ & $15$ & $3.14$ & $-122.7$ & $2.05$ & $1939\pm70$ & $4.3\pm0.2$ &  $-119.8\pm2.2$ & $2.9\pm2.3$ & $-186\pm257$ & $168\pm267$ & $-0.10\pm0.14$ & $0.09\pm0.14$ \\
         & $7$ & $14$ & $2.11$ & $-120.0$ & $1.37$ & $2091\pm38$ & $4.7\pm0.1$ &  $-122.0\pm1.0$ & $2.0\pm1.0$ & $202\pm222$ & $247\pm218$ & $0.10\pm0.11$ & $0.12\pm0.11$ \\
         & $8$ & $13$ & $0.97$ & $-119.8$ & $0.63$ & $1506\pm46$ & $3.4\pm0.1$ &  $-114.8\pm1.2$ & $5.0\pm1.3$ & $-234\pm302$ & $145\pm191$ & $-0.16\pm0.21$ & $0.10\pm0.13$ \\
         & $11$ & $11$ & $8.73$ & $-117.0$ & $5.69$ & $2109\pm31$ & $4.7\pm0.1$ &  $-116.8\pm0.7$ & $0.3\pm0.7$ & $349\pm62$ & $28\pm50$ & $0.17\pm0.03$ & $0.01\pm0.02$ \\
         & $12$ & $12$ & $7.08$ & $-116.7$ & $4.61$ & $2224\pm70$ & $5.0\pm0.2$ &  $-116.0\pm1.5$ & $0.7\pm1.7$ & $255\pm107$ & $-3\pm96$ & $0.12\pm0.05$ & $-0.00\pm0.04$ \\
         & $24$ & $11$ & $2.90$ & $-111.2$ & $1.89$ & $1906\pm40$ & $4.2\pm0.1$ &  $-105.7\pm0.7$ & $5.5\pm0.7$ & $-179\pm156$ & $-17\pm96$ & $-0.10\pm0.08$ & $-0.01\pm0.05$ \\
0458-020 & $2$ & $10$ & $4.64$ & $-55.6$ & $38.65$ & $185\pm9$ & $16.3\pm0.8$ &  $-55.8\pm2.9$ & $0.2\pm2.9$ & $-36\pm6$ & $3\pm6$ & $-0.65\pm0.11$ & $0.05\pm0.11$ \\
         & $3$ & $10$ & $2.19$ & $-45.5$ & $18.24$ & $152\pm9$ & $13.4\pm0.8$ &  $-58.2\pm3.2$ & $12.8\pm3.3$ & $-5\pm6$ & $2\pm6$ & $-0.11\pm0.13$ & $0.03\pm0.12$ \\
0528+134 & $2$ & $18$ & $1.90$ & $35.4$ & $16.05$ & $227\pm5$ & $18.9\pm0.4$ &  $21.6\pm1.0$ & $13.8\pm1.1$ & $18\pm4$ & $-4\pm3$ & $0.25\pm0.05$ & $-0.05\pm0.04$ \\
         & $3$ & $22$ & $1.16$ & $42.2$ & $9.80$ & $142\pm10$ & $11.8\pm0.8$ &  $27.6\pm3.0$ & $14.6\pm3.4$ & $-19\pm6$ & $-1\pm5$ & $-0.42\pm0.14$ & $-0.03\pm0.10$ \\
0605-085 & $2$ & $14$ & $4.74$ & $113.4$ & $36.68$ & $254\pm23$ & $12.0\pm1.1$ &  $118.8\pm3.3$ & $5.4\pm3.3$ & $17\pm14$ & $-6\pm9$ & $0.13\pm0.10$ & $-0.04\pm0.07$ \\
         & $3$ & $15$ & $2.78$ & $122.5$ & $21.51$ & $419\pm13$ & $19.8\pm0.6$ &  $122.2\pm1.3$ & $0.3\pm1.5$ & $28\pm7$ & $-2\pm5$ & $0.12\pm0.03$ & $-0.01\pm0.02$ \\
         & $4$ & $12$ & $1.42$ & $123.4$ & $10.99$ & $355\pm9$ & $16.7\pm0.4$ &  $123.3\pm1.2$ & $0.2\pm1.4$ & $8\pm11$ & $-6\pm9$ & $0.04\pm0.06$ & $-0.03\pm0.05$ \\
0716+714 & $5$ & $10$ & $0.97$ & $18.9$ & $4.39$ & $521\pm18$ & $10.1\pm0.3$ &  $16.5\pm1.3$ & $2.4\pm1.6$ & $194\pm42$ & $-76\pm28$ & $0.49\pm0.11$ & $-0.19\pm0.07$ \\
0736+017 & $1$ & $12$ & $11.16$ & $-84.7$ & $35.18$ & $358\pm9$ & $4.4\pm0.1$ &  $-58.2\pm2.0$ & $26.4\pm2.0$ & $-41\pm7$ & $69\pm9$ & $-0.14\pm0.02$ & $0.23\pm0.03$ \\
         & $5$ & $11$ & $5.06$ & $-73.4$ & $15.95$ & $1180\pm77$ & $14.6\pm0.9$ &  $-77.8\pm1.8$ & $4.4\pm2.1$ & $71\pm69$ & $-49\pm33$ & $0.07\pm0.07$ & $-0.05\pm0.03$ \\
0738+313 & $2$ & $16$ & $4.83$ & $155.2$ & $33.00$ & $246\pm10$ & $8.9\pm0.4$ &  $126.1\pm2.4$ & $29.1\pm2.5$ & $-18\pm5$ & $7\pm6$ & $-0.12\pm0.04$ & $0.05\pm0.04$ \\
         & $3$ & $17$ & $3.88$ & $161.9$ & $26.51$ & $169\pm5$ & $6.2\pm0.2$ &  $115.8\pm1.2$ & $46.0\pm1.3$ & $1\pm3$ & $-5\pm2$ & $0.01\pm0.02$ & $-0.05\pm0.02$ \\
         & $6$ & $17$ & $3.34$ & $178.2$ & $22.82$ & $96\pm3$ & $3.5\pm0.1$ &  $156.9\pm1.4$ & $21.3\pm1.4$ & $-5\pm2$ & $-6\pm1$ & $-0.08\pm0.03$ & $-0.11\pm0.02$ \\
         & $8$ & $14$ & $1.30$ & $-175.5$ & $8.88$ & $153\pm6$ & $5.6\pm0.2$ &  $-174.8\pm0.9$ & $0.7\pm1.0$ & $-1\pm4$ & $-2\pm2$ & $-0.01\pm0.05$ & $-0.02\pm0.02$ \\
0748+126 & $2$ & $11$ & $3.93$ & $117.4$ & $30.59$ & $383\pm17$ & $18.3\pm0.8$ &  $120.7\pm2.6$ & $3.3\pm2.7$ & $-13\pm17$ & $-19\pm16$ & $-0.06\pm0.08$ & $-0.09\pm0.08$ \\
         & $3$ & $11$ & $2.38$ & $113.2$ & $18.53$ & $216\pm13$ & $10.3\pm0.6$ &  $107.9\pm2.7$ & $5.3\pm2.8$ & $-19\pm10$ & $20\pm9$ & $-0.17\pm0.09$ & $0.17\pm0.08$ \\
0814+425 & $2$ & $15$ & $1.42$ & $88.7$ & $5.42$ & $53\pm3$ & $0.8\pm0.1$ &  $90.0\pm2.6$ & $1.3\pm2.6$ & $-6\pm2$ & $8\pm2$ & $-0.13\pm0.05$ & $0.20\pm0.04$ \\
0823+033 & $7$ & $11$ & $1.65$ & $28.2$ & $10.10$ & $592\pm27$ & $17.9\pm0.8$ &  $27.3\pm2.1$ & $0.9\pm2.4$ & $13\pm39$ & $-49\pm30$ & $0.03\pm0.10$ & $-0.13\pm0.08$ \\
0827+243 & $2$ & $19$ & $1.97$ & $120.9$ & $15.59$ & $407\pm13$ & $20.3\pm0.7$ &  $140.9\pm1.6$ & $20.0\pm1.7$ & $7\pm16$ & $51\pm15$ & $0.03\pm0.08$ & $0.24\pm0.07$ \\
         & $3$ & $13$ & $0.88$ & $112.8$ & $6.96$ & $394\pm25$ & $19.7\pm1.2$ &  $115.5\pm4.0$ & $2.7\pm4.5$ & $51\pm36$ & $36\pm38$ & $0.25\pm0.18$ & $0.18\pm0.19$ \\
0829+046 & $5$ & $15$ & $2.35$ & $69.5$ & $6.87$ & $903\pm34$ & $10.2\pm0.4$ &  $67.1\pm2.0$ & $2.4\pm2.2$ & $189\pm53$ & $-127\pm48$ & $0.25\pm0.07$ & $-0.17\pm0.06$ \\
         & $6$ & $10$ & $1.84$ & $67.1$ & $5.38$ & $788\pm30$ & $8.9\pm0.3$ &  $73.0\pm1.1$ & $5.9\pm1.2$ & $-519\pm104$ & $-47\pm52$ & $-0.77\pm0.16$ & $-0.07\pm0.08$ \\
0836+710 & $1$ & $13$ & $12.97$ & $-149.6$ & $108.54$ & $222\pm18$ & $19.2\pm1.6$ &  $-159.2\pm2.2$ & $9.6\pm2.2$ & $-18\pm13$ & $2\pm6$ & $-0.26\pm0.19$ & $0.03\pm0.09$ \\
         & $2$ & $11$ & $8.45$ & $-142.6$ & $70.71$ & $256\pm26$ & $22.2\pm2.2$ &  $-150.1\pm4.7$ & $7.6\pm4.7$ & $-23\pm19$ & $36\pm16$ & $-0.29\pm0.24$ & $0.46\pm0.20$ \\
         & $3$ & $12$ & $3.44$ & $-145.8$ & $28.79$ & $156\pm9$ & $13.5\pm0.8$ &  $-151.5\pm2.6$ & $5.8\pm2.6$ & $-8\pm7$ & $2\pm6$ & $-0.16\pm0.15$ & $0.04\pm0.12$ \\
         & $4$ & $12$ & $2.33$ & $-142.9$ & $19.50$ & $192\pm7$ & $16.6\pm0.6$ &  $-150.3\pm1.8$ & $7.4\pm1.9$ & $-5\pm6$ & $-0\pm5$ & $-0.08\pm0.10$ & $-0.00\pm0.08$ \\
         & $5$ & $12$ & $1.41$ & $-138.8$ & $11.80$ & $224\pm8$ & $19.4\pm0.7$ &  $-146.0\pm2.4$ & $7.2\pm2.5$ & $-1\pm6$ & $-18\pm7$ & $-0.01\pm0.09$ & $-0.26\pm0.10$ \\
         & $6$ & $12$ & $0.77$ & $-139.6$ & $6.44$ & $203\pm5$ & $17.6\pm0.4$ &  $-138.2\pm1.5$ & $1.4\pm1.8$ & $35\pm5$ & $-6\pm5$ & $0.56\pm0.08$ & $-0.09\pm0.07$ \\
0838+133 & $2$ & $14$ & $3.10$ & $79.2$ & $21.90$ & $250\pm9$ & $9.7\pm0.3$ &  $94.3\pm2.4$ & $15.1\pm2.4$ & $-8\pm9$ & $30\pm10$ & $-0.05\pm0.06$ & $0.20\pm0.07$ \\
\,(3C\,207) & $3$ & $12$ & $2.14$ & $74.6$ & $15.12$ & $334\pm30$ & $12.9\pm1.2$ &  $69.9\pm4.6$ & $4.7\pm4.6$ & $-96\pm101$ & $-233\pm82$ & $-0.48\pm0.51$ & $-1.17\pm0.43$ \\
         & $6$ & $13$ & $0.55$ & $82.9$ & $3.89$ & $54\pm7$ & $2.1\pm0.3$ &  $62.7\pm4.7$ & $20.2\pm4.7$ & $-17\pm27$ & $-32\pm16$ & $-0.52\pm0.83$ & $-0.99\pm0.51$ \\
0851+202 & $1$ & $13$ & $1.00$ & $-93.7$ & $4.48$ & $613\pm26$ & $11.8\pm0.5$ &  $-93.5\pm0.7$ & $0.2\pm0.8$ & $-222\pm64$ & $-55\pm18$ & $-0.47\pm0.14$ & $-0.12\pm0.04$ \\
\,(OJ287) & $4$ & $10$ & $1.48$ & $-116.2$ & $6.64$ & $795\pm20$ & $15.3\pm0.4$ &  $-104.9\pm2.3$ & $11.3\pm2.4$ & $60\pm120$ & $65\pm190$ & $0.10\pm0.20$ & $0.11\pm0.31$ \\
         & $5$ & $19$ & $1.18$ & $-115.9$ & $5.29$ & $540\pm6$ & $10.4\pm0.1$ &  $-102.7\pm0.4$ & $13.2\pm0.4$ & $-132\pm22$ & $125\pm13$ & $-0.32\pm0.05$ & $0.30\pm0.03$ \\
         & $9$ & $13$ & $1.12$ & $-121.3$ & $5.02$ & $462\pm38$ & $8.9\pm0.7$ &  $-110.5\pm3.1$ & $10.8\pm3.2$ & $-245\pm84$ & $46\pm55$ & $-0.69\pm0.24$ & $0.13\pm0.16$ \\
         & $15$ & $24$ & $3.35$ & $-109.3$ & $15.02$ & $273\pm21$ & $5.2\pm0.4$ &  $-111.1\pm4.5$ & $1.8\pm4.5$ & $-9\pm32$ & $33\pm32$ & $-0.05\pm0.15$ & $0.16\pm0.15$ \\
0917+624 & $3$ & $12$ & $2.66$ & $-13.1$ & $22.67$ & $229\pm17$ & $15.4\pm1.1$ &  $-13.0\pm3.0$ & $0.1\pm3.1$ & $-74\pm19$ & $-26\pm13$ & $-0.79\pm0.21$ & $-0.28\pm0.14$ \\
         & $4$ & $10$ & $1.41$ & $-22.0$ & $12.02$ & $179\pm24$ & $12.1\pm1.6$ &  $-8.8\pm4.7$ & $13.2\pm4.8$ & $-21\pm36$ & $-31\pm20$ & $-0.28\pm0.49$ & $-0.42\pm0.28$ \\
         & $5$ & $11$ & $0.89$ & $-31.7$ & $7.59$ & $135\pm8$ & $9.1\pm0.5$ &  $-15.3\pm2.8$ & $16.4\pm2.9$ & $-5\pm10$ & $-36\pm8$ & $-0.08\pm0.18$ & $-0.66\pm0.15$ \\
0945+408 & $1$ & $12$ & $11.55$ & $113.4$ & $96.98$ & $200\pm18$ & $12.2\pm1.1$ &  $113.3\pm4.8$ & $0.1\pm4.9$ & $-15\pm9$ & $7\pm10$ & $-0.17\pm0.10$ & $0.08\pm0.11$ \\
         & $4$ & $11$ & $1.23$ & $114.0$ & $10.33$ & $217\pm11$ & $13.3\pm0.7$ &  $114.6\pm1.7$ & $0.6\pm1.9$ & $16\pm7$ & $-5\pm4$ & $0.17\pm0.07$ & $-0.05\pm0.05$ \\
1055+018 & $3$ & $23$ & $3.91$ & $-49.8$ & $30.45$ & $229\pm21$ & $11.0\pm1.0$ &  $-87.3\pm4.8$ & $37.5\pm4.9$ & $-13\pm20$ & $23\pm18$ & $-0.11\pm0.16$ & $0.19\pm0.15$ \\
1127-145 & $4$ & $13$ & $4.16$ & $80.9$ & $34.65$ & $239\pm10$ & $14.1\pm0.6$ &  $84.0\pm1.9$ & $3.1\pm2.0$ & $6\pm7$ & $8\pm5$ & $0.05\pm0.06$ & $0.07\pm0.05$ \\
1150+812 & $2$ & $10$ & $3.02$ & $170.2$ & $25.36$ & $105\pm5$ & $6.4\pm0.3$ &  $155.3\pm3.7$ & $15.0\pm3.7$ & $-1\pm4$ & $-1\pm6$ & $-0.03\pm0.08$ & $-0.03\pm0.12$ \\
         & $3$ & $10$ & $2.62$ & $179.8$ & $22.00$ & $115\pm5$ & $7.0\pm0.3$ &  $172.0\pm3.2$ & $7.9\pm3.2$ & $-5\pm4$ & $-11\pm4$ & $-0.10\pm0.08$ & $-0.21\pm0.08$ \\
         & $4$ & $10$ & $1.40$ & $-160.6$ & $11.76$ & $111\pm5$ & $6.8\pm0.3$ &  $-171.3\pm1.3$ & $10.7\pm1.3$ & $9\pm4$ & $-19\pm2$ & $0.19\pm0.08$ & $-0.38\pm0.04$ \\
1156+295 & $4$ & $10$ & $0.71$ & $-3.3$ & $5.16$ & $327\pm27$ & $13.4\pm1.1$ &  $6.9\pm1.1$ & $10.2\pm1.2$ & $55\pm59$ & $-132\pm14$ & $0.29\pm0.31$ & $-0.70\pm0.09$ \\
1222+216 & $1$ & $16$ & $12.43$ & $-7.8$ & $69.59$ & $339\pm13$ & $8.9\pm0.3$ &  $7.7\pm3.0$ & $15.6\pm3.0$ & $-4\pm10$ & $17\pm13$ & $-0.02\pm0.04$ & $0.07\pm0.06$ \\
         & $2$ & $16$ & $7.15$ & $6.6$ & $40.03$ & $575\pm16$ & $15.1\pm0.4$ &  $12.7\pm1.4$ & $6.1\pm1.4$ & $-13\pm12$ & $36\pm10$ & $-0.03\pm0.03$ & $0.09\pm0.03$ \\
         & $3$ & $13$ & $3.66$ & $2.7$ & $20.49$ & $636\pm7$ & $16.7\pm0.2$ &  $9.7\pm0.3$ & $7.0\pm0.4$ & $-2\pm6$ & $26\pm3$ & $-0.00\pm0.01$ & $0.06\pm0.01$ \\
         & $4$ & $16$ & $3.25$ & $-0.9$ & $18.20$ & $633\pm7$ & $16.6\pm0.2$ &  $5.0\pm0.5$ & $5.9\pm0.6$ & $8\pm5$ & $32\pm4$ & $0.02\pm0.01$ & $0.07\pm0.01$ \\
         & $5$ & $11$ & $3.24$ & $-5.9$ & $18.14$ & $643\pm13$ & $16.9\pm0.3$ &  $0.0\pm0.4$ & $5.9\pm0.5$ & $-27\pm14$ & $27\pm5$ & $-0.06\pm0.03$ & $0.06\pm0.01$ \\
1226+023 & $1$ & $14$ & $18.69$ & $-123.7$ & $50.50$ & $820\pm53$ & $8.5\pm0.6$ &  $-143.5\pm3.4$ & $19.8\pm3.4$ & $215\pm112$ & $66\pm100$ & $0.30\pm0.16$ & $0.09\pm0.14$ \\
\,(3C\,273) & $2$ & $50$ & $13.63$ & $-122.1$ & $36.83$ & $727\pm10$ & $7.5\pm0.1$ &  $-128.5\pm0.8$ & $6.4\pm0.8$ & $52\pm6$ & $-34\pm6$ & $0.08\pm0.01$ & $-0.05\pm0.01$ \\
         & $3$ & $19$ & $6.33$ & $-113.3$ & $17.10$ & $1046\pm17$ & $10.8\pm0.2$ &  $-117.8\pm0.9$ & $4.5\pm1.0$ & $-78\pm33$ & $-8\pm36$ & $-0.09\pm0.04$ & $-0.01\pm0.04$ \\
         & $5$ & $53$ & $8.69$ & $-117.8$ & $23.48$ & $1099\pm7$ & $11.3\pm0.1$ &  $-118.9\pm0.3$ & $1.1\pm0.3$ & $-77\pm4$ & $-36\pm4$ & $-0.08\pm0.00$ & $-0.04\pm0.00$ \\
         & $7$ & $13$ & $2.03$ & $-119.8$ & $5.49$ & $1318\pm42$ & $13.6\pm0.4$ &  $-119.5\pm1.4$ & $0.3\pm1.8$ & $-35\pm123$ & $-39\pm97$ & $-0.03\pm0.11$ & $-0.03\pm0.08$ \\
         & $8$ & $14$ & $3.04$ & $-131.0$ & $8.21$ & $836\pm24$ & $8.6\pm0.2$ &  $-134.6\pm1.6$ & $3.6\pm1.9$ & $86\pm36$ & $34\pm36$ & $0.12\pm0.05$ & $0.05\pm0.05$ \\
         & $9$ & $44$ & $4.62$ & $-127.5$ & $12.48$ & $900\pm7$ & $9.3\pm0.1$ &  $-127.9\pm0.4$ & $0.4\pm0.5$ & $126\pm5$ & $18\pm5$ & $0.16\pm0.01$ & $0.02\pm0.01$ \\
         & $10$ & $16$ & $1.12$ & $-120.2$ & $3.03$ & $492\pm22$ & $5.1\pm0.2$ &  $-118.6\pm2.1$ & $1.6\pm2.4$ & $230\pm33$ & $17\pm26$ & $0.54\pm0.08$ & $0.04\pm0.06$ \\
         & $12$ & $31$ & $2.81$ & $-135.2$ & $7.59$ & $939\pm18$ & $9.7\pm0.2$ &  $-135.0\pm1.1$ & $0.2\pm1.3$ & $30\pm27$ & $52\pm27$ & $0.04\pm0.03$ & $0.06\pm0.03$ \\
         & $15$ & $17$ & $2.11$ & $-141.0$ & $5.70$ & $907\pm27$ & $9.4\pm0.3$ &  $-140.1\pm1.7$ & $0.9\pm1.8$ & $206\pm58$ & $-30\pm59$ & $0.26\pm0.07$ & $-0.04\pm0.07$ \\
1253-055 & $1$ & $89$ & $4.64$ & $-122.1$ & $29.28$ & $359\pm2$ & $11.4\pm0.1$ &  $-138.7\pm0.3$ & $16.7\pm0.4$ & $12\pm1$ & $-2\pm1$ & $0.05\pm0.01$ & $-0.01\pm0.01$ \\
\,(3C\,279) & $2$ & $34$ & $3.34$ & $-118.3$ & $21.08$ & $270\pm12$ & $8.6\pm0.4$ &  $-113.2\pm1.9$ & $5.1\pm1.9$ & $-24\pm25$ & $28\pm19$ & $-0.14\pm0.14$ & $0.16\pm0.11$ \\
         & $3$ & $12$ & $0.93$ & $-126.4$ & $5.87$ & $220\pm19$ & $7.0\pm0.6$ &  $-136.8\pm4.8$ & $10.3\pm4.8$ & $74\pm36$ & $-101\pm36$ & $0.52\pm0.25$ & $-0.70\pm0.26$ \\
         & $5$ & $76$ & $2.44$ & $-125.8$ & $15.40$ & $508\pm9$ & $16.1\pm0.3$ &  $-126.5\pm1.0$ & $0.7\pm1.1$ & $-34\pm7$ & $-11\pm7$ & $-0.10\pm0.02$ & $-0.03\pm0.02$ \\
         & $6$ & $44$ & $1.48$ & $-127.2$ & $9.34$ & $486\pm10$ & $15.4\pm0.3$ &  $-127.8\pm1.2$ & $0.6\pm1.2$ & $366\pm20$ & $41\pm20$ & $1.16\pm0.07$ & $0.13\pm0.06$ \\
         & $7$ & $28$ & $1.42$ & $-125.5$ & $8.96$ & $651\pm25$ & $20.6\pm0.8$ &  $-120.8\pm1.9$ & $4.8\pm2.0$ & $27\pm64$ & $48\pm56$ & $0.06\pm0.15$ & $0.11\pm0.13$ \\
         & $8$ & $15$ & $0.91$ & $-135.0$ & $5.74$ & $289\pm18$ & $9.2\pm0.6$ &  $-111.1\pm3.1$ & $23.9\pm3.2$ & $361\pm67$ & $197\pm60$ & $1.92\pm0.37$ & $1.05\pm0.32$ \\
         & $10$ & $11$ & $0.78$ & $-140.9$ & $4.92$ & $475\pm15$ & $15.1\pm0.5$ &  $-136.7\pm1.7$ & $4.2\pm1.9$ & $-97\pm37$ & $23\pm37$ & $-0.31\pm0.12$ & $0.07\pm0.12$ \\
1308+326 & $2$ & $37$ & $2.96$ & $-73.0$ & $23.79$ & $399\pm13$ & $20.8\pm0.7$ &  $-71.5\pm1.1$ & $1.5\pm1.2$ & $-8\pm7$ & $10\pm5$ & $-0.04\pm0.04$ & $0.05\pm0.02$ \\
         & $3$ & $12$ & $1.70$ & $-63.7$ & $13.66$ & $443\pm48$ & $23.1\pm2.5$ &  $-75.8\pm3.5$ & $12.1\pm3.6$ & $128\pm74$ & $-86\pm45$ & $0.58\pm0.34$ & $-0.39\pm0.21$ \\
         & $4$ & $21$ & $1.85$ & $-51.7$ & $14.87$ & $519\pm22$ & $27.1\pm1.1$ &  $-64.6\pm1.5$ & $12.8\pm1.6$ & $-39\pm40$ & $18\pm24$ & $-0.15\pm0.15$ & $0.07\pm0.09$ \\
         & $5$ & $34$ & $2.09$ & $-42.1$ & $16.80$ & $449\pm6$ & $23.4\pm0.3$ &  $-46.2\pm0.8$ & $4.1\pm0.9$ & $-65\pm6$ & $-17\pm6$ & $-0.29\pm0.03$ & $-0.08\pm0.03$ \\
         & $7$ & $13$ & $1.65$ & $-61.6$ & $13.26$ & $428\pm18$ & $22.3\pm0.9$ &  $-64.9\pm2.5$ & $3.3\pm2.6$ & $316\pm55$ & $-46\pm57$ & $1.47\pm0.26$ & $-0.22\pm0.26$ \\
         & $9$ & $19$ & $0.45$ & $-34.3$ & $3.62$ & $188\pm5$ & $9.8\pm0.3$ &  $-43.4\pm1.5$ & $9.1\pm1.6$ & $36\pm8$ & $-11\pm7$ & $0.38\pm0.08$ & $-0.11\pm0.08$ \\
1510-089 & $1$ & $18$ & $3.22$ & $-25.4$ & $16.10$ & $713\pm17$ & $15.9\pm0.4$ &  $-25.7\pm0.8$ & $0.3\pm0.9$ & $-95\pm22$ & $-3\pm13$ & $-0.18\pm0.04$ & $-0.01\pm0.03$ \\
         & $4$ & $10$ & $2.04$ & $-34.9$ & $10.20$ & $909\pm52$ & $20.3\pm1.2$ &  $-33.7\pm3.6$ & $1.2\pm4.0$ & $88\pm89$ & $-11\pm100$ & $0.13\pm0.13$ & $-0.02\pm0.15$ \\
1546+027 & $3$ & $12$ & $2.49$ & $172.4$ & $13.59$ & $406\pm27$ & $10.3\pm0.7$ &  $169.1\pm1.2$ & $3.4\pm1.3$ & $2\pm27$ & $-18\pm6$ & $0.01\pm0.09$ & $-0.06\pm0.02$ \\
         & $4$ & $11$ & $1.12$ & $172.5$ & $6.11$ & $171\pm20$ & $4.3\pm0.5$ &  $165.6\pm3.2$ & $6.9\pm3.4$ & $-10\pm21$ & $-14\pm9$ & $-0.08\pm0.17$ & $-0.11\pm0.08$ \\
1611+343 & $4$ & $23$ & $2.91$ & $172.4$ & $24.74$ & $64\pm5$ & $4.2\pm0.3$ &  $108.3\pm3.0$ & $64.1\pm3.0$ & $4\pm3$ & $2\pm2$ & $0.14\pm0.10$ & $0.06\pm0.08$ \\
         & $8$ & $13$ & $0.78$ & $152.8$ & $6.63$ & $212\pm9$ & $14.0\pm0.6$ &  $164.0\pm1.5$ & $11.2\pm1.8$ & $4\pm10$ & $10\pm6$ & $0.04\pm0.11$ & $0.11\pm0.06$ \\
1633+382 & $2$ & $28$ & $2.51$ & $-91.8$ & $21.43$ & $171\pm7$ & $13.2\pm0.5$ &  $-99.6\pm1.1$ & $7.8\pm1.1$ & $-32\pm5$ & $-12\pm2$ & $-0.53\pm0.08$ & $-0.21\pm0.04$ \\
         & $4$ & $13$ & $1.14$ & $-90.2$ & $9.73$ & $376\pm21$ & $29.1\pm1.6$ &  $-89.5\pm1.4$ & $0.7\pm1.6$ & $-55\pm29$ & $14\pm14$ & $-0.41\pm0.22$ & $0.11\pm0.11$ \\
         & $5$ & $13$ & $0.86$ & $-79.6$ & $7.34$ & $288\pm14$ & $22.3\pm1.1$ &  $-83.4\pm2.0$ & $3.8\pm2.2$ & $-121\pm26$ & $2\pm19$ & $-1.18\pm0.26$ & $0.02\pm0.18$ \\
         & $7$ & $15$ & $0.57$ & $-67.5$ & $4.87$ & $84\pm11$ & $6.5\pm0.8$ &  $-67.5\pm3.7$ & $0.0\pm3.9$ & $-5\pm11$ & $20\pm5$ & $-0.17\pm0.36$ & $0.68\pm0.20$ \\
1637+574 & $1$ & $12$ & $4.31$ & $-159.3$ & $31.67$ & $212\pm9$ & $8.9\pm0.4$ &  $-149.2\pm2.6$ & $10.1\pm2.6$ & $-29\pm8$ & $12\pm8$ & $-0.24\pm0.07$ & $0.10\pm0.07$ \\
1641+399 & $2$ & $28$ & $5.49$ & $-77.3$ & $36.44$ & $239\pm14$ & $8.3\pm0.5$ &  $-57.9\pm3.6$ & $19.4\pm3.6$ & $-34\pm9$ & $11\pm9$ & $-0.23\pm0.06$ & $0.07\pm0.06$ \\
\,(3C\,345) & $3$ & $32$ & $3.33$ & $-90.8$ & $22.10$ & $320\pm10$ & $11.1\pm0.3$ &  $-73.5\pm1.4$ & $17.3\pm1.4$ & $4\pm6$ & $44\pm5$ & $0.02\pm0.03$ & $0.22\pm0.02$ \\
         & $6$ & $34$ & $2.47$ & $-90.9$ & $16.39$ & $342\pm5$ & $11.8\pm0.2$ &  $-84.7\pm0.6$ & $6.2\pm0.6$ & $-1\pm3$ & $9\pm2$ & $-0.01\pm0.02$ & $0.04\pm0.01$ \\
         & $7$ & $35$ & $2.30$ & $-97.1$ & $15.26$ & $374\pm4$ & $12.9\pm0.1$ &  $-92.1\pm0.9$ & $5.0\pm1.0$ & $6\pm3$ & $9\pm4$ & $0.02\pm0.01$ & $0.04\pm0.02$ \\
         & $8$ & $30$ & $1.68$ & $-88.8$ & $11.15$ & $325\pm4$ & $11.2\pm0.1$ &  $-87.3\pm0.8$ & $1.5\pm0.9$ & $19\pm3$ & $-11\pm3$ & $0.09\pm0.01$ & $-0.05\pm0.02$ \\
         & $9$ & $27$ & $1.53$ & $-93.7$ & $10.15$ & $348\pm9$ & $12.0\pm0.3$ &  $-98.8\pm0.8$ & $5.1\pm0.8$ & $4\pm7$ & $-13\pm4$ & $0.02\pm0.03$ & $-0.06\pm0.02$ \\
         & $11$ & $18$ & $0.98$ & $-86.8$ & $6.50$ & $559\pm15$ & $19.3\pm0.5$ &  $-87.8\pm0.7$ & $1.0\pm0.8$ & $73\pm35$ & $-16\pm19$ & $0.21\pm0.10$ & $-0.05\pm0.05$ \\
         & $12$ & $15$ & $1.10$ & $-88.8$ & $7.30$ & $471\pm27$ & $16.3\pm0.9$ &  $-86.1\pm1.4$ & $2.7\pm1.7$ & $-149\pm57$ & $31\pm31$ & $-0.50\pm0.20$ & $0.10\pm0.10$ \\
         & $13$ & $11$ & $0.78$ & $-94.1$ & $5.18$ & $331\pm19$ & $11.4\pm0.7$ &  $-94.7\pm3.6$ & $0.6\pm4.1$ & $-155\pm57$ & $128\pm72$ & $-0.75\pm0.28$ & $0.62\pm0.35$ \\
1655+077 & $6$ & $11$ & $1.83$ & $-34.4$ & $12.41$ & $312\pm12$ & $11.2\pm0.4$ &  $-40.8\pm2.1$ & $6.4\pm2.4$ & $14\pm9$ & $-5\pm9$ & $0.08\pm0.05$ & $-0.03\pm0.05$ \\
1730-130 & $2$ & $16$ & $6.89$ & $14.6$ & $53.87$ & $466\pm27$ & $22.5\pm1.3$ &  $15.6\pm2.4$ & $1.0\pm2.5$ & $-37\pm15$ & $-11\pm11$ & $-0.15\pm0.06$ & $-0.04\pm0.05$ \\
         & $6$ & $15$ & $1.86$ & $27.3$ & $14.54$ & $324\pm10$ & $15.7\pm0.5$ &  $25.6\pm1.2$ & $1.7\pm1.4$ & $13\pm7$ & $-8\pm5$ & $0.08\pm0.04$ & $-0.05\pm0.03$ \\
1749+096 & $5$ & $22$ & $0.41$ & $-2.2$ & $1.90$ & $121\pm18$ & $2.4\pm0.4$ &  $22.2\pm4.2$ & $24.4\pm4.3$ & $140\pm36$ & $21\pm18$ & $1.53\pm0.45$ & $0.23\pm0.20$ \\
1800+440 & $3$ & $11$ & $4.44$ & $-162.5$ & $31.01$ & $407\pm13$ & $15.4\pm0.5$ &  $-160.6\pm2.4$ & $1.9\pm2.5$ & $-5\pm9$ & $1\pm11$ & $-0.02\pm0.04$ & $0.01\pm0.05$ \\
1823+568 & $1$ & $14$ & $7.60$ & $-160.9$ & $53.11$ & $248\pm51$ & $9.4\pm1.9$ &  $-170.8\pm4.9$ & $9.9\pm4.9$ & $34\pm58$ & $-6\pm24$ & $0.23\pm0.39$ & $-0.04\pm0.16$ \\
         & $4$ & $28$ & $3.90$ & $-158.7$ & $27.25$ & $550\pm13$ & $20.9\pm0.5$ &  $-162.6\pm1.2$ & $3.9\pm1.3$ & $-16\pm12$ & $-24\pm8$ & $-0.05\pm0.04$ & $-0.07\pm0.03$ \\
         & $6$ & $13$ & $1.84$ & $-161.8$ & $12.86$ & $152\pm11$ & $5.8\pm0.4$ &  $-164.8\pm3.4$ & $3.0\pm3.4$ & $-40\pm26$ & $52\pm20$ & $-0.44\pm0.29$ & $0.57\pm0.22$ \\
         & $12$ & $22$ & $1.97$ & $-161.4$ & $13.77$ & $88\pm7$ & $3.3\pm0.3$ &  $-153.7\pm4.6$ & $7.7\pm4.7$ & $-7\pm10$ & $-25\pm9$ & $-0.14\pm0.19$ & $-0.46\pm0.18$ \\
1828+487 & $2$ & $12$ & $11.86$ & $-34.8$ & $84.34$ & $322\pm9$ & $12.6\pm0.3$ &  $-37.9\pm1.3$ & $3.1\pm1.3$ & $-4\pm7$ & $-1\pm6$ & $-0.02\pm0.04$ & $-0.00\pm0.03$ \\
\,(3C\,380) & $3$ & $12$ & $10.75$ & $-31.8$ & $76.45$ & $266\pm3$ & $10.4\pm0.1$ &  $-38.2\pm0.7$ & $6.4\pm0.7$ & $4\pm2$ & $-13\pm3$ & $0.02\pm0.01$ & $-0.08\pm0.02$ \\
         & $5$ & $12$ & $5.12$ & $-28.4$ & $36.41$ & $348\pm10$ & $13.7\pm0.4$ &  $-26.1\pm1.1$ & $2.2\pm1.1$ & $13\pm7$ & $5\pm5$ & $0.06\pm0.03$ & $0.02\pm0.02$ \\
         & $6$ & $11$ & $2.50$ & $-29.3$ & $17.78$ & $321\pm11$ & $12.6\pm0.4$ &  $-31.8\pm1.5$ & $2.5\pm1.6$ & $-16\pm9$ & $-5\pm7$ & $-0.08\pm0.05$ & $-0.03\pm0.04$ \\
         & $8$ & $10$ & $2.05$ & $-38.4$ & $14.58$ & $310\pm6$ & $12.2\pm0.2$ &  $-38.1\pm0.9$ & $0.2\pm0.9$ & $18\pm4$ & $3\pm4$ & $0.10\pm0.02$ & $0.02\pm0.02$ \\
1928+738 & $1$ & $32$ & $11.70$ & $166.8$ & $51.98$ & $447\pm18$ & $8.5\pm0.3$ &  $-180.0\pm0.9$ & $13.2\pm0.9$ & $-11\pm10$ & $3\pm4$ & $-0.03\pm0.03$ & $0.01\pm0.01$ \\
\,(4C\,+73.18) & $2$ & $12$ & $6.32$ & $165.2$ & $28.08$ & $384\pm58$ & $7.3\pm1.1$ &  $172.7\pm3.8$ & $7.5\pm3.8$ & $-63\pm68$ & $-4\pm31$ & $-0.21\pm0.23$ & $-0.02\pm0.11$ \\
         & $3$ & $32$ & $4.56$ & $163.5$ & $20.26$ & $227\pm11$ & $4.3\pm0.2$ &  $174.0\pm1.8$ & $10.5\pm1.8$ & $9\pm6$ & $1\pm4$ & $0.05\pm0.03$ & $0.00\pm0.02$ \\
         & $4$ & $10$ & $3.56$ & $170.8$ & $15.81$ & $328\pm27$ & $6.2\pm0.5$ &  $173.8\pm1.8$ & $3.0\pm1.8$ & $113\pm66$ & $31\pm28$ & $0.45\pm0.26$ & $0.12\pm0.11$ \\
         & $5$ & $30$ & $3.12$ & $170.4$ & $13.86$ & $281\pm12$ & $5.3\pm0.2$ &  $167.4\pm0.9$ & $2.9\pm0.9$ & $-11\pm7$ & $-7\pm3$ & $-0.05\pm0.03$ & $-0.03\pm0.01$ \\
         & $6$ & $28$ & $2.64$ & $159.7$ & $11.73$ & $256\pm4$ & $4.9\pm0.1$ &  $171.8\pm0.7$ & $12.1\pm0.8$ & $-15\pm3$ & $28\pm2$ & $-0.08\pm0.01$ & $0.14\pm0.01$ \\
         & $8$ & $32$ & $1.82$ & $157.4$ & $8.09$ & $271\pm4$ & $5.2\pm0.1$ &  $163.5\pm0.4$ & $6.1\pm0.5$ & $7\pm2$ & $5\pm1$ & $0.04\pm0.01$ & $0.02\pm0.01$ \\
         & $9$ & $28$ & $1.42$ & $151.5$ & $6.31$ & $246\pm8$ & $4.7\pm0.1$ &  $151.3\pm1.3$ & $0.2\pm1.5$ & $4\pm5$ & $2\pm4$ & $0.02\pm0.03$ & $0.01\pm0.02$ \\
         & $10$ & $23$ & $1.02$ & $156.7$ & $4.53$ & $212\pm7$ & $4.0\pm0.1$ &  $153.2\pm1.7$ & $3.6\pm1.9$ & $24\pm5$ & $-3\pm5$ & $0.15\pm0.03$ & $-0.02\pm0.03$ \\
         & $11$ & $22$ & $0.64$ & $157.9$ & $2.84$ & $134\pm4$ & $2.5\pm0.1$ &  $156.1\pm1.4$ & $1.8\pm1.7$ & $-8\pm4$ & $3\pm3$ & $-0.08\pm0.03$ & $0.03\pm0.03$ \\
1957+405 & $3$ & $21$ & $2.32$ & $-79.4$ & $2.49$ & $57\pm9$ & $0.2\pm0.0$ &  $-90.0\pm3.0$ & $10.6\pm3.0$ & $-22\pm5$ & $-4\pm2$ & $-0.40\pm0.12$ & $-0.08\pm0.03$ \\
\,(Cygnus A) & $4$ & $21$ & $1.54$ & $-81.8$ & $1.66$ & $55\pm9$ & $0.2\pm0.0$ &  $-95.5\pm3.9$ & $13.7\pm4.0$ & $-13\pm4$ & $-3\pm2$ & $-0.24\pm0.09$ & $-0.06\pm0.04$ \\
         & $6$ & $21$ & $0.89$ & $-82.4$ & $0.96$ & $45\pm5$ & $0.2\pm0.0$ &  $-100.1\pm4.0$ & $17.7\pm4.1$ & $1\pm3$ & $1\pm2$ & $0.02\pm0.08$ & $0.03\pm0.04$ \\
2021+614 & $4$ & $18$ & $0.58$ & $24.0$ & $2.09$ & $29\pm3$ & $0.4\pm0.0$ &  $0.0\pm3.6$ & $24.0\pm3.7$ & $-8\pm2$ & $-6\pm1$ & $-0.32\pm0.10$ & $-0.26\pm0.06$ \\
2121+053 & $2$ & $10$ & $1.01$ & $-69.8$ & $8.58$ & $163\pm7$ & $13.1\pm0.6$ &  $-56.4\pm2.8$ & $13.4\pm3.1$ & $18\pm5$ & $12\pm6$ & $0.33\pm0.09$ & $0.21\pm0.10$ \\
2128-123 & $3$ & $10$ & $5.52$ & $-148.4$ & $33.61$ & $228\pm5$ & $6.8\pm0.1$ &  $-160.6\pm0.6$ & $12.2\pm0.6$ & $-32\pm5$ & $14\pm2$ & $-0.21\pm0.03$ & $0.09\pm0.02$ \\
2131-021 & $1$ & $13$ & $1.56$ & $101.5$ & $13.15$ & $192\pm12$ & $12.0\pm0.8$ &  $100.0\pm3.1$ & $1.5\pm3.3$ & $0\pm9$ & $17\pm6$ & $0.00\pm0.10$ & $0.21\pm0.07$ \\
         & $3$ & $10$ & $1.58$ & $114.0$ & $13.32$ & $222\pm5$ & $13.8\pm0.3$ &  $97.0\pm3.0$ & $17.1\pm3.1$ & $6\pm9$ & $-67\pm21$ & $0.07\pm0.10$ & $-0.69\pm0.21$ \\
         & $4$ & $13$ & $0.86$ & $98.5$ & $7.25$ & $83\pm4$ & $5.2\pm0.2$ &  $78.8\pm3.1$ & $19.6\pm3.2$ & $-8\pm3$ & $2\pm5$ & $-0.23\pm0.09$ & $0.06\pm0.14$ \\
2134+004 & $2$ & $29$ & $1.74$ & $-67.7$ & $14.79$ & $73\pm5$ & $5.9\pm0.4$ &  $-145.7\pm3.6$ & $78.0\pm3.7$ & $-11\pm3$ & $7\pm3$ & $-0.44\pm0.13$ & $0.30\pm0.12$ \\
2136+141 & $3$ & $11$ & $0.83$ & $-89.5$ & $6.84$ & $59\pm2$ & $5.4\pm0.2$ &  $-132.8\pm2.3$ & $43.2\pm2.4$ & $-8\pm1$ & $-6\pm1$ & $-0.47\pm0.08$ & $-0.34\pm0.08$ \\
         & $4$ & $14$ & $0.53$ & $-71.6$ & $4.37$ & $39\pm2$ & $3.5\pm0.2$ &  $-126.1\pm2.8$ & $54.6\pm2.9$ & $-5\pm1$ & $-3\pm1$ & $-0.40\pm0.10$ & $-0.28\pm0.11$ \\
2145+067 & $2$ & $17$ & $0.74$ & $124.0$ & $5.94$ & $42\pm3$ & $2.2\pm0.2$ &  $127.0\pm4.5$ & $3.0\pm4.6$ & $1\pm2$ & $-3\pm2$ & $0.03\pm0.10$ & $-0.14\pm0.11$ \\
         & $3$ & $12$ & $0.41$ & $113.9$ & $3.29$ & $48\pm2$ & $2.5\pm0.1$ &  $146.4\pm2.6$ & $32.5\pm2.9$ & $1\pm2$ & $-2\pm2$ & $0.02\pm0.09$ & $-0.09\pm0.08$ \\
2200+420 & $1$ & $10$ & $4.76$ & $174.8$ & $6.17$ & $2341\pm163$ & $10.7\pm0.8$ &  $150.8\pm3.1$ & $24.1\pm3.2$ & $-212\pm688$ & $-343\pm540$ & $-0.10\pm0.31$ & $-0.16\pm0.25$ \\
\,(BL Lac) & $2$ & $23$ & $4.75$ & $173.1$ & $6.15$ & $1595\pm83$ & $7.3\pm0.4$ &  $155.3\pm3.0$ & $17.8\pm3.3$ & $-189\pm155$ & $-334\pm164$ & $-0.13\pm0.10$ & $-0.22\pm0.11$ \\
         & $3$ & $21$ & $3.23$ & $-168.2$ & $4.18$ & $893\pm72$ & $4.1\pm0.3$ &  $179.5\pm1.6$ & $12.3\pm1.7$ & $454\pm191$ & $-153\pm66$ & $0.54\pm0.23$ & $-0.18\pm0.08$ \\
         & $4$ & $40$ & $4.03$ & $-178.1$ & $5.22$ & $1331\pm50$ & $6.1\pm0.2$ &  $157.2\pm1.8$ & $24.7\pm1.9$ & $406\pm75$ & $-283\pm61$ & $0.33\pm0.06$ & $-0.23\pm0.05$ \\
         & $5$ & $37$ & $2.47$ & $-164.8$ & $3.20$ & $760\pm26$ & $3.5\pm0.1$ &  $-175.9\pm1.2$ & $11.1\pm1.4$ & $474\pm45$ & $-53\pm27$ & $0.67\pm0.07$ & $-0.07\pm0.04$ \\
         & $6$ & $35$ & $2.34$ & $-165.4$ & $3.03$ & $876\pm32$ & $4.0\pm0.1$ &  $-171.6\pm1.9$ & $6.2\pm2.1$ & $544\pm51$ & $-225\pm46$ & $0.66\pm0.07$ & $-0.27\pm0.06$ \\
         & $10$ & $24$ & $1.77$ & $-163.9$ & $2.29$ & $1102\pm66$ & $5.0\pm0.3$ &  $-163.9\pm1.9$ & $0.0\pm2.1$ & $442\pm149$ & $51\pm83$ & $0.43\pm0.15$ & $0.05\pm0.08$ \\
         & $11$ & $20$ & $4.01$ & $-177.0$ & $5.20$ & $2084\pm54$ & $9.5\pm0.2$ &  $162.3\pm1.7$ & $20.7\pm2.1$ & $474\pm116$ & $-620\pm133$ & $0.24\pm0.06$ & $-0.32\pm0.07$ \\
         & $12$ & $16$ & $2.97$ & $-165.8$ & $3.85$ & $1214\pm47$ & $5.5\pm0.2$ &  $-170.7\pm1.1$ & $4.8\pm1.3$ & $-39\pm118$ & $-140\pm47$ & $-0.03\pm0.10$ & $-0.12\pm0.04$ \\
         & $13$ & $14$ & $2.49$ & $-167.5$ & $3.23$ & $1098\pm33$ & $5.0\pm0.1$ &  $-168.0\pm2.0$ & $0.5\pm2.2$ & $75\pm71$ & $-201\pm73$ & $0.07\pm0.07$ & $-0.20\pm0.07$ \\
         & $16$ & $10$ & $1.65$ & $-176.7$ & $2.14$ & $684\pm82$ & $3.1\pm0.4$ &  $-166.8\pm2.3$ & $9.9\pm2.4$ & $612\pm253$ & $-10\pm88$ & $0.96\pm0.41$ & $-0.02\pm0.14$ \\
2201+315 & $1$ & $13$ & $4.68$ & $-135.9$ & $20.45$ & $332\pm10$ & $6.2\pm0.2$ &  $-129.7\pm1.5$ & $6.2\pm1.5$ & $-9\pm6$ & $-1\pm6$ & $-0.03\pm0.02$ & $-0.00\pm0.02$ \\
         & $2$ & $14$ & $3.81$ & $-142.2$ & $16.65$ & $336\pm9$ & $6.2\pm0.2$ &  $-136.2\pm1.6$ & $6.0\pm1.7$ & $-15\pm5$ & $4\pm6$ & $-0.06\pm0.02$ & $0.02\pm0.02$ \\
         & $5$ & $10$ & $1.93$ & $-149.0$ & $8.43$ & $345\pm16$ & $6.4\pm0.3$ &  $-143.4\pm2.7$ & $5.6\pm2.9$ & $41\pm16$ & $22\pm16$ & $0.15\pm0.06$ & $0.08\pm0.06$ \\
2223-052 & $1$ & $14$ & $5.29$ & $97.6$ & $44.99$ & $219\pm19$ & $14.5\pm1.3$ &  $89.2\pm3.3$ & $8.4\pm3.3$ & $16\pm22$ & $-3\pm15$ & $0.17\pm0.24$ & $-0.04\pm0.16$ \\
\,(3C\,446) & $2$ & $22$ & $3.14$ & $103.4$ & $26.71$ & $86\pm8$ & $5.7\pm0.5$ &  $94.1\pm3.5$ & $9.3\pm3.5$ & $-7\pm4$ & $6\pm2$ & $-0.20\pm0.11$ & $0.17\pm0.07$ \\
         & $4$ & $18$ & $0.73$ & $86.9$ & $6.21$ & $260\pm7$ & $17.2\pm0.5$ &  $111.6\pm1.3$ & $24.7\pm1.7$ & $37\pm6$ & $34\pm5$ & $0.34\pm0.06$ & $0.31\pm0.05$ \\
         & $5$ & $10$ & $0.59$ & $105.1$ & $5.02$ & $218\pm15$ & $14.4\pm1.0$ &  $90.7\pm3.2$ & $14.4\pm4.1$ & $25\pm16$ & $-11\pm14$ & $0.27\pm0.18$ & $-0.12\pm0.16$ \\
2230+114 & $2$ & $20$ & $10.95$ & $157.6$ & $88.83$ & $173\pm12$ & $9.3\pm0.6$ &  $137.2\pm4.0$ & $20.4\pm4.0$ & $-1\pm10$ & $-3\pm10$ & $-0.01\pm0.11$ & $-0.04\pm0.11$ \\
\,(CTA 102) & $7$ & $12$ & $2.75$ & $150.7$ & $22.31$ & $181\pm28$ & $9.7\pm1.5$ &  $175.4\pm2.8$ & $24.8\pm3.0$ & $-9\pm21$ & $25\pm7$ & $-0.10\pm0.23$ & $0.28\pm0.09$ \\
         & $9$ & $18$ & $1.17$ & $132.9$ & $9.49$ & $161\pm10$ & $8.6\pm0.5$ &  $144.3\pm3.6$ & $11.4\pm3.7$ & $-30\pm9$ & $3\pm10$ & $-0.38\pm0.12$ & $0.04\pm0.12$ \\
2243-123 & $2$ & $11$ & $3.59$ & $19.9$ & $24.54$ & $151\pm9$ & $5.5\pm0.3$ &  $37.7\pm3.6$ & $17.8\pm3.7$ & $1\pm6$ & $-0\pm7$ & $0.01\pm0.07$ & $-0.00\pm0.07$ \\
         & $3$ & $11$ & $1.86$ & $-1.4$ & $12.72$ & $121\pm5$ & $4.4\pm0.2$ &  $7.8\pm0.6$ & $9.3\pm0.6$ & $13\pm3$ & $1\pm1$ & $0.17\pm0.05$ & $0.01\pm0.01$ \\
2251+158 & $2$ & $53$ & $5.87$ & $-80.7$ & $45.21$ & $84\pm4$ & $3.9\pm0.2$ &  $-37.1\pm3.2$ & $43.7\pm3.3$ & $-9\pm3$ & $-4\pm3$ & $-0.19\pm0.06$ & $-0.10\pm0.06$ \\
\,(3C\,454.3) & $3$ & $12$ & $2.61$ & $-82.9$ & $20.10$ & $163\pm15$ & $7.6\pm0.7$ &  $-37.0\pm5.0$ & $46.0\pm5.0$ & $85\pm44$ & $-145\pm43$ & $0.97\pm0.51$ & $-1.66\pm0.51$ \\
         & $4$ & $33$ & $2.00$ & $-106.0$ & $15.40$ & $304\pm17$ & $14.2\pm0.8$ &  $-109.6\pm2.9$ & $3.5\pm3.0$ & $29\pm24$ & $-57\pm19$ & $0.18\pm0.15$ & $-0.35\pm0.12$ \\
         & $5$ & $49$ & $1.58$ & $-36.1$ & $12.17$ & $128\pm8$ & $6.0\pm0.4$ &  $-41.0\pm3.3$ & $4.9\pm3.4$ & $14\pm5$ & $-31\pm4$ & $0.21\pm0.07$ & $-0.46\pm0.07$ \\
         & $8$ & $18$ & $0.50$ & $-89.8$ & $3.85$ & $291\pm9$ & $13.6\pm0.4$ &  $-95.6\pm1.3$ & $5.7\pm1.4$ & $-350\pm41$ & $-107\pm32$ & $-2.23\pm0.27$ & $-0.68\pm0.21$ \\
         & $11$ & $34$ & $0.90$ & $-93.2$ & $6.93$ & $120\pm9$ & $5.6\pm0.4$ &  $-92.5\pm3.1$ & $0.8\pm3.1$ & $85\pm16$ & $-6\pm11$ & $1.31\pm0.26$ & $-0.09\pm0.17$ \\
2345-167 & $1$ & $12$ & $3.34$ & $120.1$ & $21.86$ & $400\pm24$ & $13.5\pm0.8$ &  $115.1\pm1.8$ & $5.0\pm2.1$ & $-8\pm16$ & $3\pm9$ & $-0.03\pm0.06$ & $0.01\pm0.03$ \\
\enddata
\tablecomments{
Columns are as follows: 
(1) Source name in B1950 coordinates (Alternate source name); 
(2) Component ID;
(3) Number of epochs;
(4) Mean radial separation from core in milli-arcseconds (averaged over all epochs);
(5) Mean structural position angle in degrees;
(6) Mean projected radial distance in parsecs;
(7) Angular proper motion in micro-arcseconds per year;
(8) Apparent speed in units of the speed of light;
(9) Proper motion position angle in degrees;
(10) Absolute difference between mean structural positon angle and proper motion position angle in degrees;
(11) Angluar acceleration parallel to the proper motion position angle in micro-arcseconds per year per year;
(12) Angular acceleration perpendicular to the proper motion position angle in micro-arcseconds per year per year;
(13) Relative parallel acceleration as defined in \S{};
(14) Relative perpendicular acceleration as defined in \S{};   
}
\end{deluxetable*} 

\end{document}